# A General Neural Network Potential for Energetic Materials with C, H, N, and O elements

**Mingjie Wen, Jiahe Han, Wenjuan Li, Xiaoya Chang, Qingzhao Chu*, Dongping Chen**

*State Key Laboratory of Explosion Science and Safety Protection, Beijing Institute of Technology, Beijing 100081, P. R. China*

**Abstract:** The discovery and optimization of high-energy materials (HEMs) are constrained by the prohibitive computational expense and prolonged development cycles inherent in conventional approaches. In this work, we develop a general neural network potential (NNP) that efficiently predicts the structural, mechanical, and decomposition properties of HEMs composed of C, H, N, and O. Our framework leverages pre-trained NNP models, fine-tuned using transfer learning on energy and force data derived from density functional theory (DFT) calculations. This strategy enables rapid adaptation across 20 different HEM systems while maintaining DFT-level accuracy, significantly reducing computational costs. A key aspect of this work is the ability of NNP model to capture the chemical activity space of HEMs, accurately describe the key atomic interactions and reaction mechanisms during thermal decomposition. The general NNP model has been applied in molecular dynamics (MD) simulations and validated with experimental data for various HEM structures. Results show that the NNP model accurately predicts the structural, mechanical, and decomposition properties of HEMs by effectively describing their chemical activity space. Compared to traditional force fields, it offers superior DFT-level accuracy and generalization across both microscopic and macroscopic properties, reducing the computational and experimental costs. This work provides an efficient strategy for the design and development of HEMs and proposes a promising framework for integrating DFT, machine learning, and experimental methods in materials research. (To facilitate further research and practical applications, we open-source our NNP model on GitHub: https://github.com/MingjieWen/General-NNP-model-for-C-H-N-O-Energetic-Materials.)

**Keywords:** High-Energy Materials (HEMs); Neural Network Potential (NNP); Transfer Learning; Chemical Activity Space; Molecular Dynamics (MD) Simulation

* Corresponding author, E-mail: chuqz@bit.edu.cn (Q. Z. Chu).

# 1 Introduction

The synthesis of high energy density energetic materials (HEMs) is a long-standing goal pursued by scientists worldwide [1, 2]. These HEMs, composed of high-energy molecular crystals formed by C, H, N, and O elements, are critical for practical industrial applications due to their unique micro- and macro-properties [1, 3]. Traditional experiment-driven material design is time-consuming and costly, relying on extensive trial-and-error and expensive setups. [4, 5]. Analyzing and mining massive amounts of experimental data requires a lot of time and manpower, with unknown patterns in the data difficult to efficiently discover using manual efforts or existing theoretical knowledge, thereby limiting the pace of research progress [6]. The rapid rise of computational materials science has shifted material development from the "experience + trial-and-error" model to a computation-driven mode [7, 8]. The computation-driven mode based on molecular dynamics (MD) simulation can significantly enhance material development efficiency and reduce costs through high-throughput calculations, batch analyses, and formula pre-screening [8, 9]. Modern physical chemistry techniques help clarify structures, reactions, and mechanical properties at the atomic levels [10, 11]. However, existing simulation methods struggle to balance accuracy and efficiency. Therefore, continued research on these materials in limited and extended systems is strongly encouraged. Developing a strategy capable of predicting the crystal structural, mechanical, and decomposition properties of HEMs while maintaining high accuracy and efficiency is particularly crucial.

Currently, the primary tools for industrial MD simulations are *ab initio* molecular dynamics (AIMD) and reactive force field molecular dynamics (ReaxFF-MD) [12, 13, 14, 15]. AIMD can accurately calculate quantum mechanical atomic forces at each step, but it struggles with high-dimensional complex systems [13, 14]. While AIMD offers high precision in predictions, its high computational cost and long simulation times pose significant challenges for the rapid iteration and optimization needed in new material development. In recent years, the reactive force field (ReaxFF), developed by van Duin et al. [16, 17, 18] has become a leader in this field, widely used in studying the decomposition and combustion processes of HEMs [12, 17, 19, 20, 21]. ReaxFF describes both reactive and non-reactive interactions between atoms using bond order-dependent polarizable charge descriptions [16, 17]. It parametrizes interatomic interactions using relatively complex functional forms. However, ReaxFF struggles to accurately describe the reaction potential energy surface (PES) with density functional theory (DFT) precision, especially when it is applied blindly to new molecules [22, 23].

Recently, machine learning (ML) technology has advanced rapidly, highlighting an increasingly vital role in materials science [24, 25]. Many ML models now aid in computational material discovery [22, 25, 26], with studies showing that incorporating configuration space information from the PES enhances chemical space exploration, balancing accuracy and efficiency [27, 28]. Neural networks are widely used for PES fitting, with notable methods including the Behler-Parrinello Neural Network (BPNN) [29], Gaussian Approximation Potential (GAP) [30], Gradient Domain Machine Learning (GDML) [26], and Deep Potential (DP scheme) [31, 32]. Among these, the DP scheme has excelled in modeling isolated molecules [31], multi-body clusters [33, 34], and solid materials [35, 36], enabling atomistic descriptions of complex reactions beyond electronic structure calculations, such as extreme physical and chemical processes [37, 38], oxidative combustion [39], and explosions [40, 41]. Our recent work extends NNP-based MD simulations to the microscopic scale [41, 42, 43], bridging electronic structure calculations, first-principles simulations, and multiscale modeling. This approach accurately describes mechanical, chemical, and thermal processes in various systems with DFT-level precision while being more efficient than traditional force fields and DFT calculations [42, 43, 44, 45], sometimes providing semi-quantitative property predictions [42, 44, 45].

Most ML projects rely on supervised learning [46], which plays a key role in computation-driven material development [8, 25, 31]. However, the need for large datasets and high computational costs limits its practicality [47]. To address this, we explore more efficient modeling approaches that maintain accuracy while reducing time and cost. Transfer learning and pre-trained models have gained attention as effective strategies for optimizing ML models [48, 49]. Transfer learning leverages existing data, reducing the need for extensive training, accelerating learning, and improving performance [48]. For example, Wang et al. [49] used the Deep Potential generator (DP-GEN) framework [50] to develop a transferable ML potential for Ag-Au nanoalloys, accurately predicting surface diffusion and nanoparticle formation even without surface data in training. In our previous work, we developed an NNP model for the three components of RDX, HMX, and CL-20 [42], capable of describing their mechanical and thermal decomposition properties. However, its scalability remained uncertain. In this work, we take a pioneering step toward a scalable and generalizable framework using a pre-trained NNP model and transfer learning, moving closer to MD simulations with chemical accuracy.

Building on the pre-trained NNP-CHNO-2024 model and a transfer learning scheme, we developed an accurate and efficient general NNP model for condensed-phase chemistry in C, H, N, O HEMs. Using the DP-GEN process, only ～20% of new training data was added, with new structures not explicitly included in the existing training database. The atomic energy and force predictions of the NNP model were thoroughly evaluated against DFT calculations. Additionally, principal component analysis (PCA) was used to explore the chemical activity space, providing deeper insights into atomic interactions and reactivity. The model was successfully applied to predict the crystal structural, mechanical, and decomposition properties of 20 different HEMs, with predictions rigorously compared to experiment reports. This work has achieved unprecedented levels of efficiency and accuracy in exploring the physicochemical space and thermal decomposition behavior of HEMs, facilitating a comprehensive assessment of their microscopic properties, guided by the insights derived from the chemical activity space.

# 2 Methodology

## 2.1 Data set preparation

Here, a total of 20 different HEMs composed of C, H, N, and O elements are considered in the general NNP model. As shown in Fig. 1, these HEMs were categorized based on their configurations into ionic salts (ADN, TKX-50, and TAGN), chain-like structures (FOX-7, NG, PETN, and BTTN), cyclic-like structures (RDX, HMX, TNT, DNBF, BTF, TATB, DTTO, NTO, TNB, HNS, NC), and cage-like structures (CL-20 and TEX), with detailed abbreviations listed in Table S1. We believe our initial configurations encompass the main C, H, N, O HEMs. Since dataset quality and size directly impact NNP model performance, we ensured comprehensive PES coverage to meet training and testing requirements. Two strategies were employed to generate the dataset for these HEMs.

### 2.1.1 AIMD calculation

The foundational DFT dataset is derived from our previous work, where AIMD simulations were performed on RDX, HMX, and CL-20 over a wide temperature range (300-4000 K) using the CP2K package [51]. Each HEM in the system was simulated at temperatures of 300, 1000, 2000, 3000, and 4000 K, with a simulation time of 2 ps for each temperature. This resulted in obtaining 5000 snapshots of energy and force data for each HEM, totaling 15000 structures. During training, 25% of the configurations were randomly selected as the test set. 10 configurations were randomly selected from the test set to calculate the training loss for each training batch. For specific computational details, please refer to our previous work [42, 43, 52, 53].

### 2.1.2 DP-GEN scheme

As shown in Fig. 1, based on the foundational DFT dataset, the DP-GEN process [50] from the DeePMD-kit package [54] was used to sample and generate new configurations during the transfer learning process. In the transfer learning process, four NNPs were obtained by training with different initialization seeds. One of these NNPs was used to perform four MD simulations over the temperature range of 300 to 4000 K. The other three NNPs were used to evaluate the MD trajectories and obtain deviations in atomic forces, serving as bounds for identifying new configurations. Configurations on the MD trajectories were recorded at time intervals of $\Delta t$ = 0.02 ps.

**Table 1** Configuration dataset and training scheme of general NNP model.

| Structure | Iteration | Scaling factors | Temperature(K) | Sampling No. (AIMD/DP-GEN) [a] |
|---|---|---|---|---|
| RDX | | 0.92, 0.96, | | 5000/4000 |
| HMX | - | 1.00, 1.04, | 300-4000 | 5000/4000 |
| CL-20 | | 1.08 | | 5000/4000 |
| TNT | 1-5 | 1.00 | 300-4000 | 0/800 |
| ADN | 5-9 | 1.00 | 300-4000 | 0/400 |
| FOX-7 | 10-11 | 1.00 | 300-4000 | 0/400 |
| TKX-50 | 12-13 | 1.00 | 300-4000 | 0/400 |
| DNBF | 14-16 | 1.00 | 300-4000 | 0/300 |
| BTF | 17-19 | 1.00 | 300-4000 | 0/300 |
| TATB | 20-21 | 1.00 | 300-4000 | 0/200 |
| TAGN | 22-23 | 1.00 | 300-4000 | 0/200 |
| NG | 24-25 | 1.00 | 300-4000 | 0/200 |
| PETN | 26-27 | 1.00 | 300-4000 | 0/500 |
| DTTO/iso-DTTO | 28-30 | 1.00 | 300-4000 | 0/200 |
| NTO | 31-33 | 1.00 | 300-4000 | 0/200 |
| TEX | 34-35 | 1.00 | 300-4000 | 0/200 |
| BTTN | 36-38 | 1.00 | 300-4000 | 0/500 |
| NC | 39-40 | 1.00 | 300-4000 | 0/500 |
| TNB | 41-42 | 1.00 | 300-4000 | 0/800 |
| HNS | 43-45 | 1.00 | 300-4000 | 0/800 |

[a] The AIMD simulations and DP-GEN processes on the foundational DFT dataset were conducted across various temperatures (300 K - 4000 K). Structural coordinates were transformed according to the scaling factors in the x, y, and z directions.

The explored configurations are divided into three categories according to the error indicators: accurate, candidate, and failed. Accurate configurations matched well with the DFT results, while failed configurations had significant errors. Of course, configurations between these extremes were labeled as candidates. In this work, the lower ($\varepsilon_{lo}$) and upper ($\varepsilon_{up}$) bounds of the force deviation were set at $\varepsilon_{lo}$ = 0.05 eV/Å and $\varepsilon_{hi}$ = 0.15 eV/Å, respectively. These screening bounds are derived from the test results of Zhang et al. [50]. Configurations with a relative force deviation of less than 0.05 eV/Å were marked as "accurate", while those within the 0.05 - 0.15 eV/Å range were marked as "candidates" and included in the training set.

The entire DP-GEN process performed 45 iterations and 6900 configurations, significantly reducing the DFT computational cost by using model deviation as an error measure. This approach distinctly differs from traditional high-cost methods that stack large amounts of data. Configuration space was sampled through NVT simulations over a wide temperature range (300 K to 4000 K).

## 2.2 General NNP model training

Fig. 1 shows the training process of the general NNP model developed using the DP scheme. Similar to our previous work [42, 45, 52, 53], the NNP model training process used physical parameters as atomic coordinates(*R*), energy (*E*), and force (*F*) as inputs for deep neural network training. The DP model represents the total potential energy (*E*) of the system as the sum of atomic energies ($E_i$) ($E = \Sigma_i E_i$). Here, each atomic energy

($E_i$) is parameterized using the NNP model, defined as,

$$E_{\mathrm{i}} = E^{\omega\alpha_{\mathrm{i}}}(r^i) \tag{1}$$

where, $r^i$ is the local environment of atom $i$ relative to its neighboring atoms within the cutoff radius in Cartesian coordinates, $\alpha_i$ is the chemical species of atom $i$, and $\omega\alpha_i$ denotes the optimized set of NNP parameters during the training process. Each subnetwork of $E_i$ consists of an embedded neural network and a fitting neural network. Both networks use the ResNet architecture. The embedding network is sized (25, 50, 100), with an embedding matrix size of 12. The fitting network is sized (240, 240, 240). The atomic cutoff radius is set to 6.0 Å, and the descriptor decays smoothly from 0.5 to 6.0 Å. The NNP model is initialized with random numbers, and each iteration involves 500,000 training steps. The initial learning rate is set to 0.001 with a decay rate of 0.98, decayed every 4000 steps. The training process iteratively adjusts the model parameters to minimize the loss function ($L$), which measures the difference between the neural network prediction and the reference data. The loss function is defined as the sum of the squared differences of the NNP predictions,

$$L = \frac{p_{\mathrm{e}}}{N}\Delta E^2 + \frac{p_{\mathrm{f}}}{3N}\sum_i \left|\Delta F_i\right|^2 \tag{2}$$

where, $p_e$ and $p_f$ represent the weights of the energy and force terms, respectively. $N$ represents the number of atoms in the structure. The DP scheme trains the model by computing gradients of the loss function using backpropagation. The NNP model is trained for $4.0 \times 10^6$ iterations, with the learning rate following an exponential decay. The entire computational process used an NVIDIA V100 GPU with 8 CPU cores.

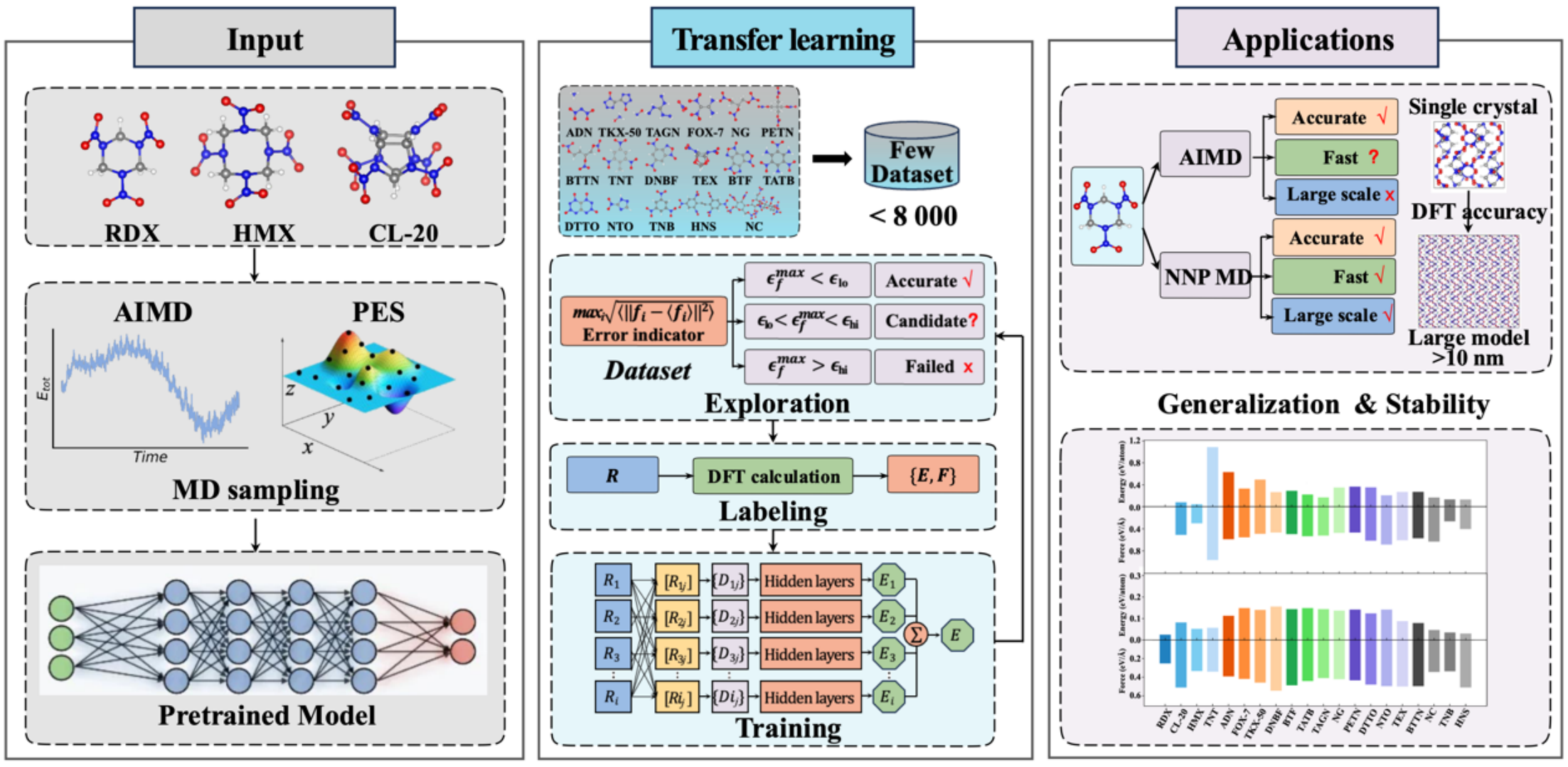

**Fig. 1** Transfer learning strategy for NNP model.

## 2.3 NNP Testing

### 2.3.1 Static Testing

We conducted crystal cell parameters and equation of state tests for all HEMs using DFT, NNP, and ReaxFF methods. Initial configurations for all crystals were sourced from the CCDC database [55], except for NG. In the DFT calculations, core electrons were treated with the Goedecker-Teter-Hutter (GTH) Gaussian-type pseudopotential using the Perdew-Burke-Ernzerhof (PBE) generalized gradient approximation methods [56, 57]. Dispersion interactions were accounted for using Grimme's DFT-D3 method [58], with energy cutoffs set to 60 Ry for wave functions and 400 Ry for electron density. Polarized orbitals were included alongside double-ζ Gaussian basis functions (DZVP-MOLOPT). SCF self-consistent field calculations were converged to an accuracy of $1.0 \times 10^{-6}$.

Additionally, the ReaxFF was adapted from Liu et al. [16]. Both NNP and ReaxFF methods were implemented in the LAMMPS package [59], with the $v_{max}$ parameter set to allow maximum box size changes of 0.001 Å per relaxation step. Convergence criteria for minimization were set at $1.0 \times 10^{-7}$ for energy and $1.0 \times 10^{-15}$ for forces.

### 2.3.2 MD simulations

To assess the performance of the general NNP model in the thermal decomposition of HEMs, MD simulations were conducted using the LAMMPS program [59]. Initially, due to varying molecular weights, HEMs were extended via supercell operations to ensure that each thermal decomposition system contained more than 32 HEM molecules (> 1500 atoms). The MD simulations of these HEMs were divided into relaxation and thermal decomposition steps. Both steps were executed using the NVT ensemble with a time step of 0.2 fs, and temperature control was maintained using the Nosé-Hoover thermostat [60]. Motion equations were integrated using the velocity Verlet method, with periodic boundary conditions (PBC) applied. The relaxation simulation was performed at 10 ps, maintaining a temperature of 300 K. During the thermal decomposition simulations of HEMs, temperatures ranged from 300 to 3000 K under heating conditions of 27 K/ps, using the final snapshots from the relaxation process as the initial state for the decomposition simulations. To ensure statistical significance of the results, all simulations were independently performed three times with different random seeds under the same conditions, and the simulation time was set to 100 ps. Finally, chemical species were analyzed from MD trajectories using the ReacNetGenerator program [61], and quantities of all products were averaged across the three simulations to mitigate numerical errors.

## 3 Results and Discussion

### 3.1 Evaluation of general NNP model

#### 3.1.1 Validation

The energy and force prediction capabilities of the general NNP model were evaluated using the training dataset in Fig. 2, with a batch size of 200. In Fig. 2(a), the energy and force predictions of the NNP model for 20 HEMs align closely along the diagonal, showing excellent fitting ability. The MAE distribution range for energy compared to DFT calculations is 0.0 - 1.0 eV/atom, and for force, it is -40 - 40 eV/Å. Specifically, Fig. 2(b) shows that the MAE distribution for energy is mainly ± 0.1 eV/atom, with a few deviations, and the MAE range for force is mostly within ± 5 eV/Å. This demonstrates the strong predictive capabilities of the NNP model over a wide temperature range. In contrast, energy and force distributions from previously pre-trained models [42] (Fig. S1) show significant drift. Although they predict the energies and forces of certain HEMs like RDX, HMX, and CL-20 well, they lack good extrapolation performance for most configurations.

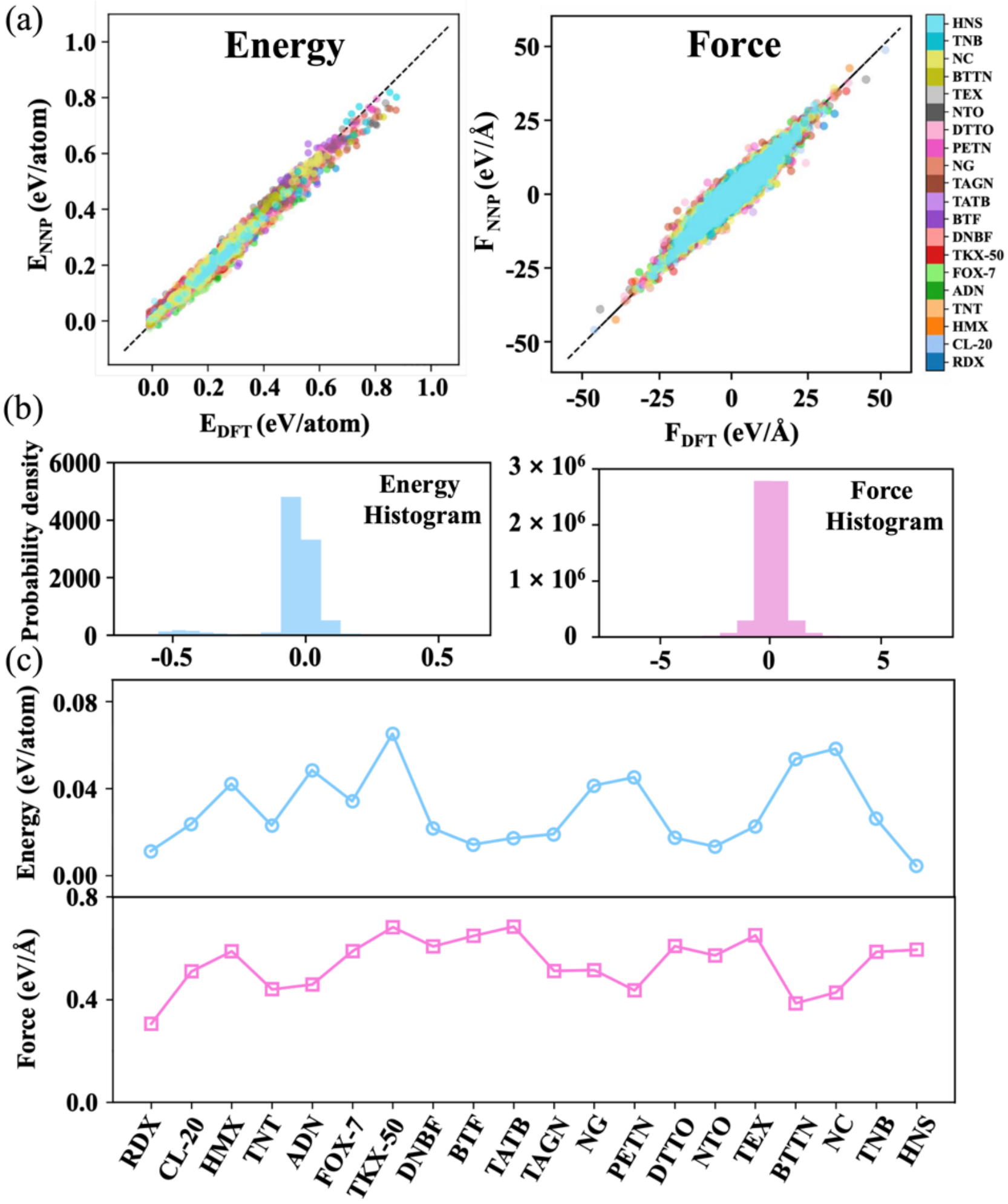

**Fig. 2** The evaluation of energy (eV/atom) and forces (eV/Å) predicted by the NNP model compared to DFT calculations (a), distribution of energy and forces in the training set (b), and MAE of energy and forces for single components (c).

Fig. 2(c) shows the energy and force predictions of the general NNP model for individual HEMs. For energy, the MAE between predicted values and DFT calculations is below 65.2 meV/atom, with the maximum observed in TKX-50. For atomic forces, the MAE is below 0.684 eV/Å, with the maximum observed in TATB. Notably, due to transfer learning from a pre-trained model [42], the general NNP predicts energies of 0.0113, 0.0422, and 0.0237 eV/atom, and forces of 0.306, 0.588, and 0.510 eV/Å for RDX, HMX, and CL-20 crystals, respectively. This accuracy is consistent across all components, demonstrating that the general NNP model can accurately describe atomic-scale decomposition reactions in both finite and extended systems of energetic materials while maintaining DFT-level precision.

### 3.1.2 Generalization of NNP Model

The excellent generalization capability of the NNP model is crucial for its practical application, and is often tested through stability and extrapolation performance. Stability refers to the ability of the NNP model to consistently predict energy and force within the data range, while extrapolation performance measures prediction accuracy beyond the training data. These characteristics directly influence the applicability and robustness of the NNP model in real simulation systems. Fig. 3 provides an in-depth evaluation of the extrapolation and stability performance of the general NNP model by comparing the MAE of energy and force predictions with DFT calculations during the training process.

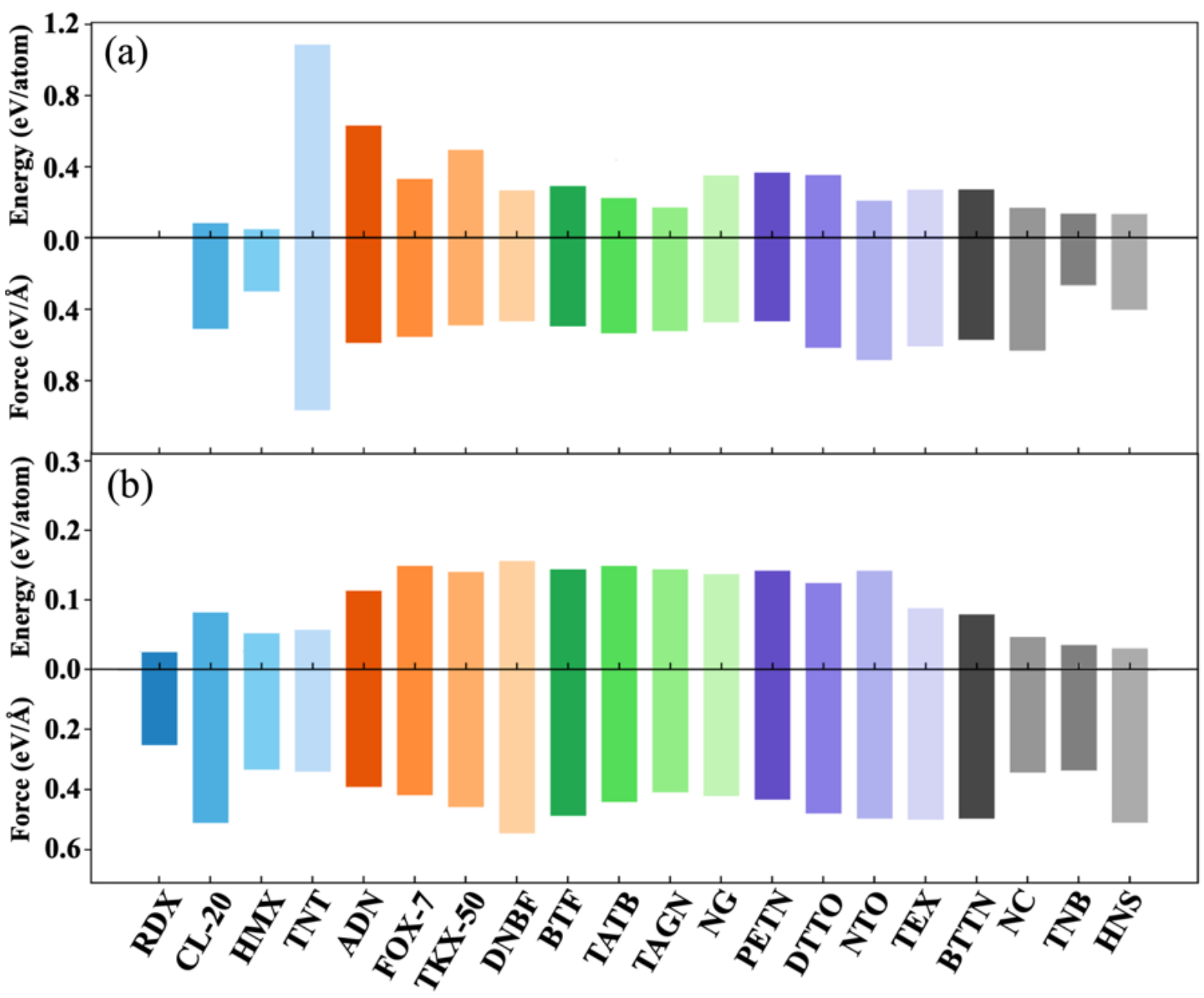

**Fig. 3** Extrapolation performance (a) and stability (b) tests of the NNP model.

As shown in Fig. 3(a), the general NNP model, initially based on pre-trained models for RDX, HMX, and CL-20, was tested for extrapolation using TNT, a HEM outside the pre-trained set. The MAE values for energy and force were 1.08 eV/atom and 0.97 eV/Å, indicating that while the model has some extrapolation ability, it cannot accurately predict all C, H, N, and O HEMs. After adding TNT to the training set, the extrapolation performance improved significantly, with energy and force MAE values for ADN dropping to 0.632 eV/atom and 0.587 eV/Å. Further inclusion of ADN in the training set reduced the MAE values for FOX-7 to 0.333 eV/atom and 0.553 eV/Å. The NNP model maintained excellent predictive performance, achieving MAE values below 0.50 eV/atom and 0.684 eV/Å for energy and force predictions. This indicates that the general NNP model, which covers common C, H, N, and O HEMs, demonstrates strong extrapolation capabilities for these extended HEMs.

Fig. 3(b) shows the stability verification results of the model, which proves that the general NNP model is highly stable in terms of energy and force prediction capabilities. Specifically, the current model shows average MAE values for energy and force for existing data compared to DFT results below 0.2 eV/atom and 0.6 eV/Å, respectively. Therefore, our general NNP model has good energy and force stability and strong extrapolation capability for C, H, N, and O HEMs in finite and extended systems, capable of revealing the microscopic properties of these HEMs at the DFT level.

### 3.2 Crystal properties

#### 3.2.1 Cell parameters

This work aims to develop a general NNP model to explain the microscopic mechanical and chemical properties of common C, H, N, and O HEMs. Crystallographic parameters were predicted using DFT, NNP, and ReaxFF methods, with results shown in Fig. 4 and Table S2. Experimental values are from the CCDC database[55], DFT calculations were performed at the PBE/DZVP-MOLOPT level using CP2K [51], and ReaxFF parameters were taken from Liu et al. [16].

As shown in Fig. 4, DFT-calculated crystal volumes closely match experimental values [55], with absolute deviations of 0%-6%, confirming the reliability of the DP-GEN training dataset. The general NNP model also shows good agreement, with deviations of 1%-16% from experiments and 1%-14% from DFT. The largest deviations occur for BTF (16% from experiments) and TKX-50 (14% from DFT). In contrast, ReaxFF predictions show much higher deviations (9%-29% from experiments and 5%-26% from DFT), with the largest errors for TAGN (29% and 26%).

The general NNP model predicts a/b/c dimensions with average absolute deviations of 1%-5% from experimental and DFT results, outperforming ReaxFF (3%–9% and 2%-8%, respectively) as shown in Table S2. However, for BTF crystallographic parameters, NNP shows larger errors than ReaxFF (5% *vs.* 4% for Cell and 16% *vs.* 14% for Volume). Despite this, NNP performs better for most HEMs, demonstrating its capability to reproduce crystallographic parameters with DFT-level accuracy.

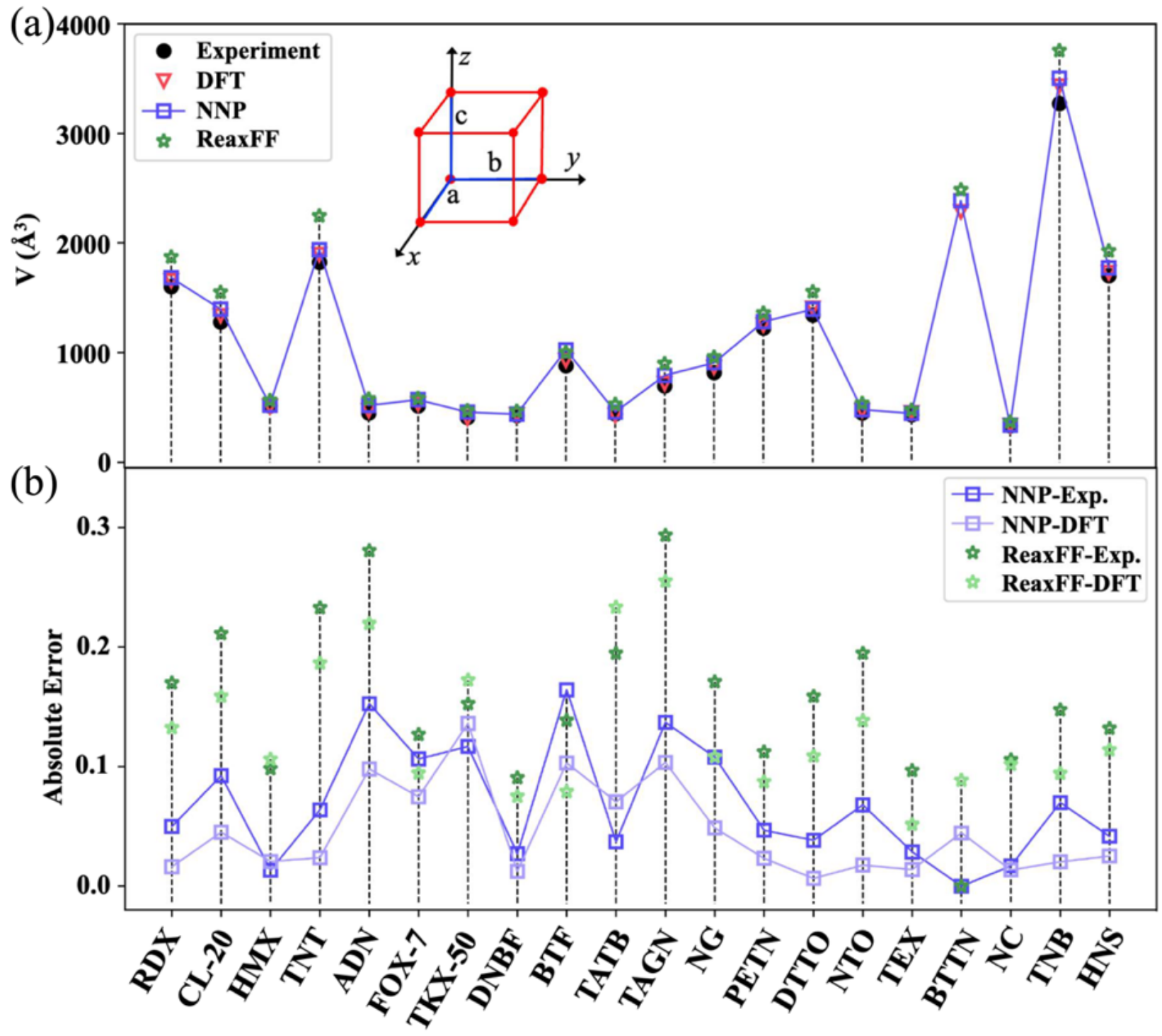

**Fig. 4** Predicted volumes of HEM crystals calculated using DFT, NNP, and ReaxFF methods and experimental results (a) and absolute errors (b).

### 3.2.2 Equation of state (EOS)

The capability of the general NNP model to reproduce the microscopic mechanical behavior of common HEMs is validated. The Equation of State (EOS) for all HEMs was obtained using DFT, NNP, and ReaxFF methods, and the mechanical behavior was extrapolated under extreme compression (0.80-0.92) and tension (1.08-1.20). In Fig. 5 and Fig. S4, shaded areas represent structures from the training set, while points outside indicate NNP predictions. Fig. 5 presents the EOS for representative ionic, chain, cyclic, and cage-like HEMs.

As shown in Fig. 5, the NNP model accurately reproduces DFT results, effectively identifying energy minima and performing well even for structures outside the training set. In contrast, ReaxFF, while predicting stable points reasonably well, shows significant deviations under tensile and compressive conditions, overestimating potential energy during compression (5-10 Å$^3$/atom). This overestimation may create artificial high-energy regions, affecting explosion and impact studies. Additional results in Fig. S4 further highlight the superior EOS prediction of the NNP model, closely aligning with DFT.

Therefore, by applying certain tensile and compressive stresses to the initial structures, considering structural deviations enhances the extrapolation potential of the NNP model, showing predictive ability even for structures beyond the training set. This capability is attributed to incorporating atomic interaction forces in the loss function, improving accuracy and capturing the microscopic mechanical behavior of HEMs.

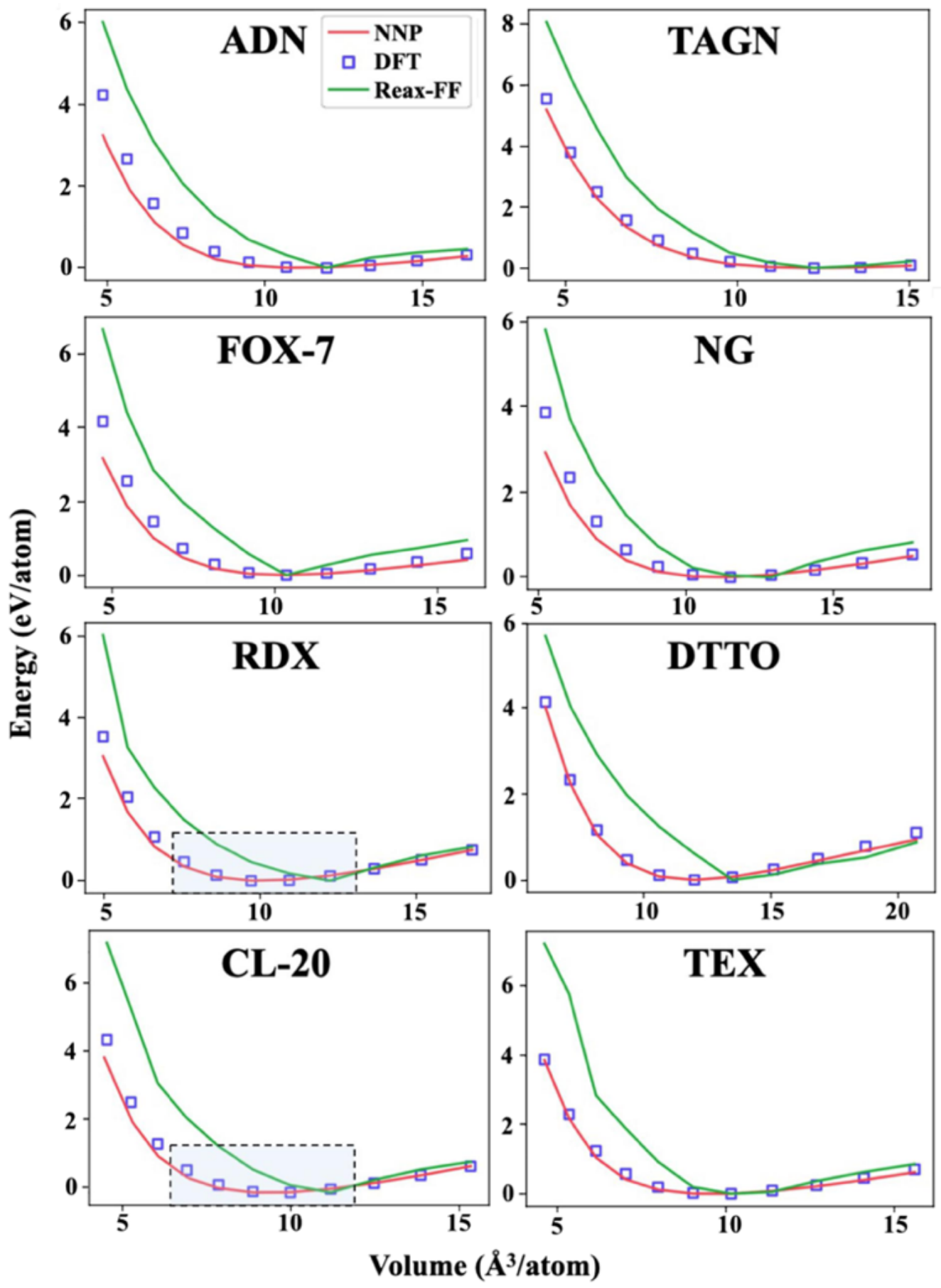

**Fig. 5** EOS curves for representative HEMs crystals: ionic salts (ADN and TAGN), chain (FOX-7 and NG), cyclic (RDX and DTTO), and cage (CL-20 and TEX) structures. "NNP" refers to the NNP model developed in this study. "DFT" calculations are computed at the PBE/DZVP-MOLOPT level using CP2K [51]. The ReaxFF is taken from the work of Liu et al. [16]. The shaded area indicates the structures included in the training set.

### 3.3 Thermal decomposition

Another goal of the general NNP model is to analyze the thermal decomposition of C, H, N, and O HEMs. MD simulations using the NNP model and ReaxFF were conducted from 300 K to 3000 K to extract major decomposition products. Fig. 6 shows the types and quantities of key products for 8 representative HEMs, including ionic salts (ADN, TAGN), chain (FOX-7, NG), cyclic (RDX, DTTO), and cage (CL-20, TEX) structures. The simulation results for all 20 HEMs are presented in Fig. S5.

As shown in Fig. 6, the main thermal decomposition products of ionic salt ADN ($NH_4N_3O_4$) include $H_2O$ (37.4%), $N_2$ (30.1%), $NO_2$ (9.7%), OH (8.7%), and NO (8.0%), closely matching ReaxFF results with slight differences in $N_2$, $NO_2$, and NO quantities. These findings align with thermogravimetry-differential thermal analysis-mass spectrometry-infrared spectroscopy (TG-DTA-MS-IR) experimental data from Izato et al. [62], which identified $H_2O$, $N_2$, $NO_2$, NO, $N_2O$, and $HNO_2$ as key gases. Similarly, TAGN ($CN_6H_9NO_3$), another ionic salt, decomposes into $CO_2$ (40.1%), $H_2O$ (30.0%), $N_2$ (12.8%), $H_2$ (4.3%), OH (2.8%), NO (2.6%), and CHNO (2.6%), with ReaxFF predictions at a similar level. Experimental studies report a decomposition temperature of ~580.28 K for TAGN[63, 64], while our model predicts 780.6 K. Notably, ReaxFF simulations show premature structural breakdown (< 500 K) in these salts, highlighting the reliability of the NNP model in predicting thermal decomposition behavior.

For chain HEMs, FOX-7 and NG serve as representatives. FOX-7 decomposes mainly into $N_2$ (32.8%), $H_2O$ (27.6%), $CO_2$ (22.8%), $CHO_2$ (3.2%), OH (3.2%), $H_2$ (3.2%), and CHNO (3.2%). The key difference from ReaxFF results lies in $CO_2$ content, as ReaxFF predicts a high presence of $C_2O_3$, whereas the NNP model aligns with AIMD simulations by Liu et al. [65], which identified $H_2O$, $CO_2$, and $N_2$ as the primary products, with $CO_2$ content close to or slightly lower than $N_2$. The NNP predictions also match experimental findings reporting $H_2O$, $N_2$, $CO_2$, NO, $NO_2$, $NH_3$, $H_2$, and CHNO [66, 67]. For NG, the NNP model predicts $CO_2$ (37.1%), $H_2O$ (28.9%), $N_2$ (14.5%), NO (6.3%), and OH (5.7%), consistent with FTIR and T-jump/Raman spectroscopy results by Hiyoshi et al. [68] and Roos et al. [69]. While the ReaxFF model predicts similar product types, it underestimates $CO_2$ and $H_2O$ due to the formation of intermediates like $HNO_2$, HNO, $HNO_3$, and $C_2H_2O_2$.

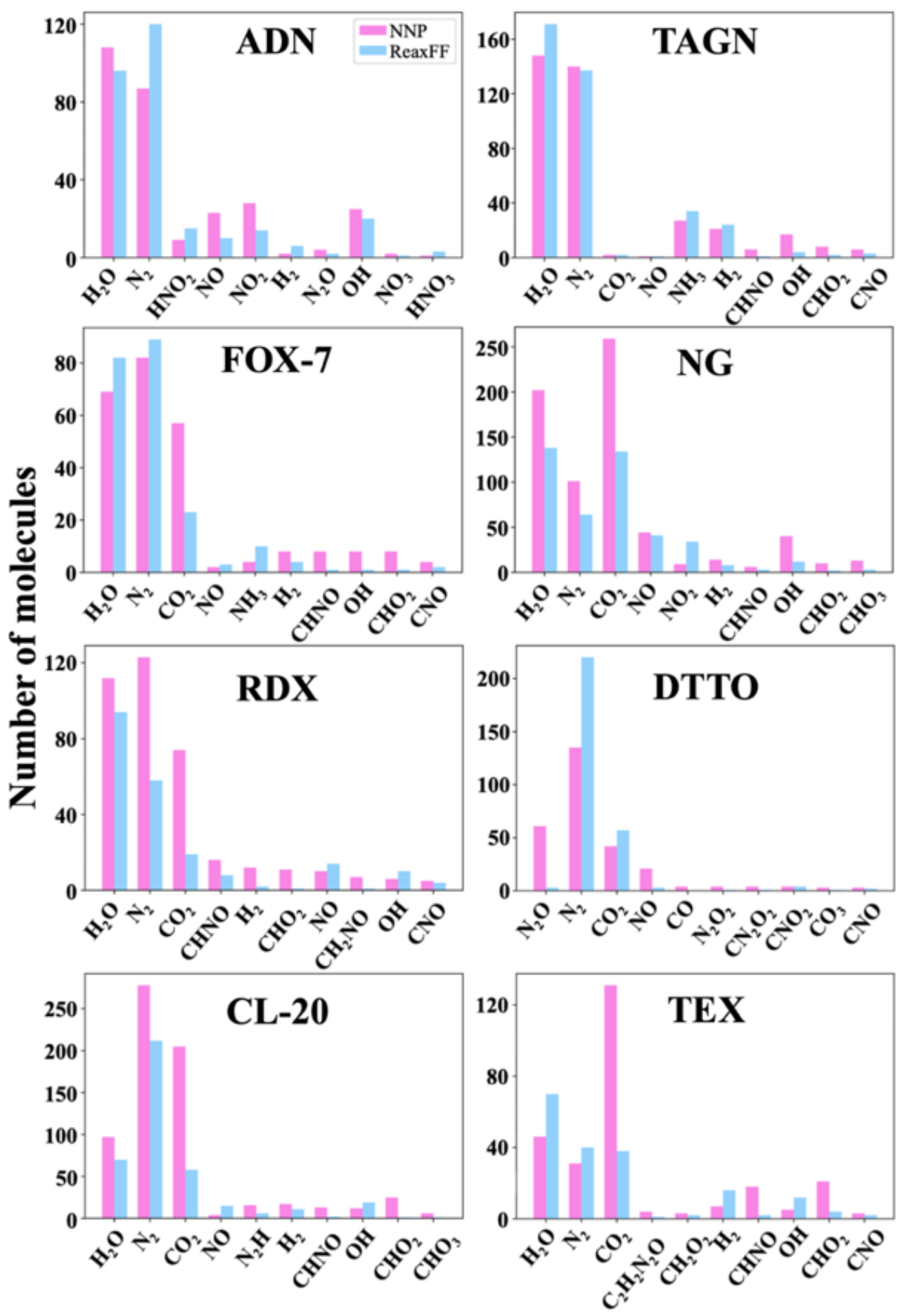

**Fig. 6** Decomposition of gaseous products of different type HEMs: ionic salts (ADN and TAGN), chain (FOX-7 and NG), cyclic (RDX and DTTO), and cage (CL-20 and TEX) structures, predicted by the NNP model (pink) and ReaxFF method (blue).

For the cyclic HEM RDX, the NNP model predicts major decomposition products as $N_2$ (32.7%), $H_2O$ (29.8%), $CO_2$ (19.7%), CHNO (4.3%), and $H_2$ (3.2%), closely matching experimental results by Ornellas et al. [70] ($N_2$: 37%, $H_2O$: 31%, $CO_2$: 18%) and findings by Khichar et al.[71] and Gongwer et al. [72], which identified $H_2O$, $N_2$, $CO_2$, CHNO, $NO_2$, and NO. For DTTO-c1, a cyclic HEM without H atoms, the NNP model predicts $N_2$ (46.8%), $N_2O$ (18.9%), $CO_2$ (16.8%), and NO (8.8%), while ReaxFF inferred more $N_2O$ active species as $N_2$. These results align with DFT-MD simulations by Ye et al. [73], which observed DTTO-c1 decomposing into two $N_2O$ molecules and DTTO-c2 forming a dimer before releasing $N_2$. This consistency highlights the reliability of the NNP model in accurately predicting cyclic HEM decomposition.

In the thermal decomposition of CL-20, the NNP model predicts major products as $N_2$ (43.3%), $CO_2$ (30.5%), $H_2O$ (14.4%), and $CHO_2$ (3.7%), matching well with Naik et al. [74] confirmed the presence of $N_2$, $CO_2$, $N_2O$, and NO among the decomposition products using thermal decomposition GC/MS studies. Additionally, Isayev et al. [75] noted significant differences in their AIMD simulations of CL-20 compared to ReaxFF simulations, particularly in the concentrations and reaction rates of $H_2O$, CO, and $CO_2$, which is also confirmed in our work. For TEX, the NNP model predicts $CO_2$ (48.7%), $H_2O$ (17.1%), $N_2$ (11.5%), $CHO_2$ (7.8%), and CHNO (6.7%), consistent with AIMD simulations by Dong et al. [76] and Zuo et al. [77] infrared spectroscopy results. In contrast, the ReaxFF model underestimates $CO_2$ for both CL-20 and TEX, likely due to the presence of long-chain C-N heterocycles ($C_2$-$C_{15}$) in ReaxFF simulations, which hinder full decomposition of TEX.

Thus, the general NNP model for C, H, N, and O HEMs not only consistent well with DFT results in describing the thermal decomposition behaviors but also quantitatively predicts experimental observations. This work breaks down barriers with experimental observations, reduces the computational complexity associated with traditional electronic structure methods, and narrows the gap between the efficiency of classical force fields and the accuracy of DFT methods.

### 3.4 Formation mechanism of chemical activity space

To elucidate the intrinsic relationship between the NNP model and the chemical activity space of the 20 HEMs, Principal Component Analysis (PCA) was applied using the Sklearn method to reduce the complex high-dimensional data of C, H, N, and O into a simpler low-dimensional visualization. This transformation reflects the inherent similarities within the original dataset. The atomic local environment descriptor was generated using the Smooth Overlap of Atomic Positions (SOAP) method with a cutoff radius of 5.0 Å. The HEMs were classified based on their molecular configurations/types. Fig. 7 shows the PCA results during training, where each point represents a configuration from the training set, colored by the 20 HEM compounds. Data density is indicated by the point color, and 800 configurations were randomly selected for plotting.

The results in Fig. 7 demonstrate that the PCA-based visualization of the dataset effectively classifies different substances in chemical space based on structures. This highlights the advantages of the SOAP descriptor in extracting chemical information, surpassing the limitations of relying solely on raw structural data. As shown in Fig. 7(a), RDX, HMX, and CL-20 exhibit distinct clustering trends in the high-dimensional space. After introducing substances with different structures, such as TNT, ADN, and FOX-7, into the pre-trained model, the chemical activity space of the general NNP model is largely established. With the iterative training of the model, new HEMs gradually fill this space, expanding the coverage of the training set. The trained NNP model is able to effectively map 20 different HEMs to the known high-dimensional chemical space, ensuring accurate predictions and generalization capability for HEMs and structurally similar substances within the training set.

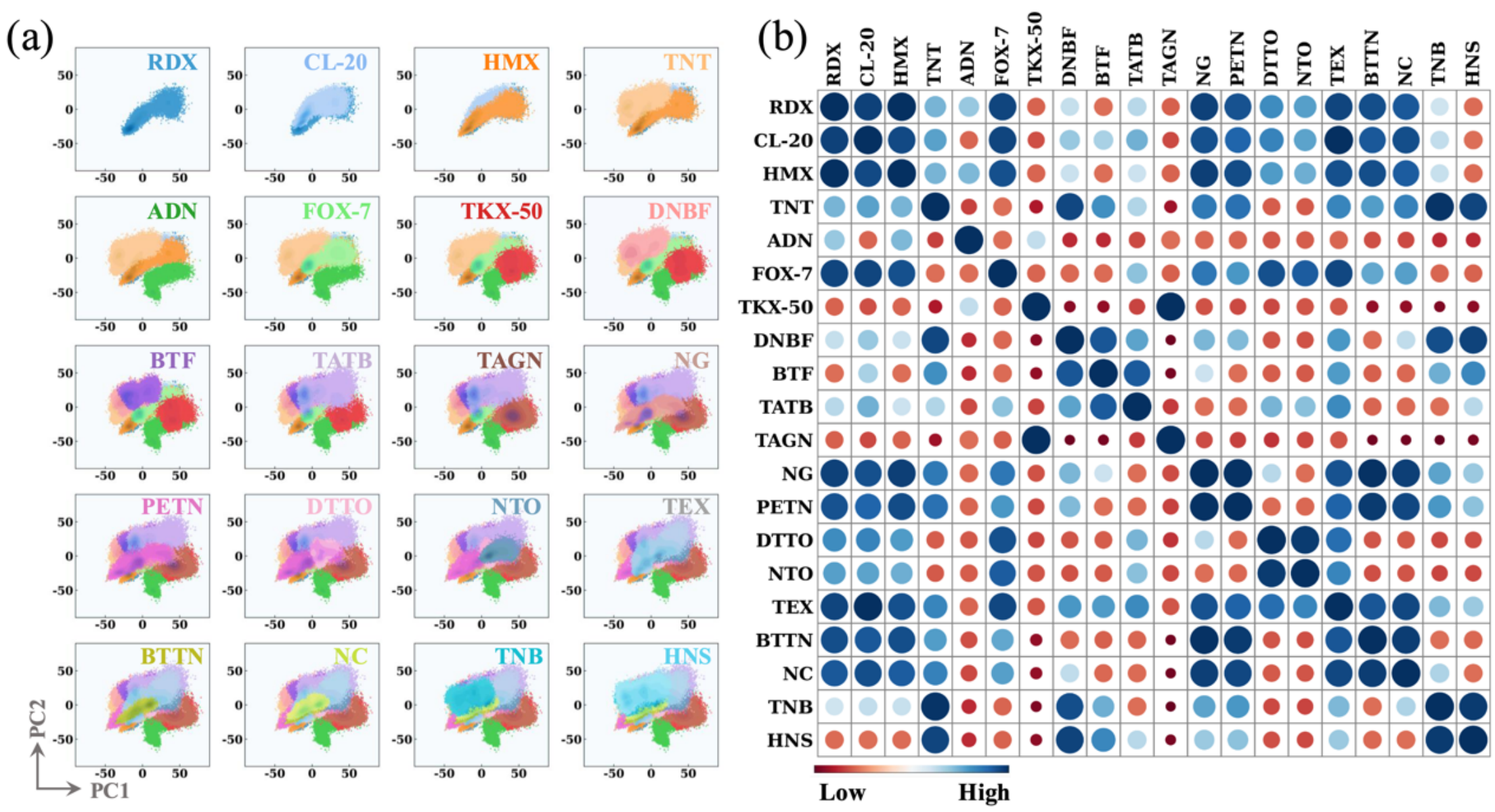

**Fig. 7** PCA visualization results of the single component and configuration space of the 20 HEMs dataset during the NNP model training (a) and the sample correlation heatmap (b).

Fig. 7(b) presents the sample correlation heatmap of the 20 HEMs, derived based on Euclidean distance, providing a quantitative evaluation of the similarity between different HEM structural features. The results indicate that the clustering trend of RDX, HMX, and CL-20 in chemical space aligns closely with their high correlation in chemical structure. Simultaneously, the data correlation heatmap shows a lower correlation between RDX, HMX, and CL-20 with other HEMs such as TNT, ADN, TKX-50, DNBF, BTF, TATB, TAGN, TNB, and HNS, particularly for ionic salts structures of ADN, TKX-50, and TAGN, confirming their significant separation in chemical space. Further analysis reveals that although some HEMs (e.g., FOX-7, NG, PETN, DTTO, NTO, TEX, BTTN, NC, etc.) differ structurally from RDX, HMX, and CL-20, they still maintain a high correlation, especially those containing $NO_2$ groups. This suggests that despite structural differences, these materials may exhibit similar physical properties or reaction mechanisms under specific chemical environments or reaction conditions, thus revealing the potential predictive power of the model.

Additionally, Figs. S6-S8 display the PCA results and sample correlation heatmaps of the 20 HEMs configuration spaces under the conditions of 300 K, 1500 K, and 3000 K during the NNP model training, corresponding to the initial, pyrolysis, and oxidation stages of HEMs. The results indicate that at 300 K, the HEMs are in the solid state, with distinct separation in their chemical activity space due to differences in crystal structure, exhibiting low correlation and insufficient formation of an effective chemical activity space. At 1500 K, during the pyrolysis stage, these HEMs partially decompose into intermediate products and small molecular gas products, forming a preliminary chemical activity space, with reduced spatial separation and increased correlation (Fig. S7(b) *vs.* Fig. S6(b)). At 3000 K, during the oxidation stage of small molecular gas products, the HEMs transition into small molecular gas products, with their spatial distribution exhibiting a high degree of similarity, rapidly forming the chemical activity space, and further increasing correlation. Materials with similar known decomposition patterns show higher correlation, while HEMs with distinct configurations exhibit lower correlation with the main compounds.

In the chemical activity space, HEMs are grouped into four categories based on their structure: C-N ring-containing structures (RDX, HMX, CL-20, NTO, DTTO), phenyl ring-containing structures (TNT, DNBF, HNS, TNB, TATB), chain-like structures (NG, BTTN, PETN, FOX-7), and ionic salts (TAGN, TKX-50, ADN). PCA analysis shows clear clustering of these compounds in high-dimensional space (Fig. 8). RDX, HMX, and CL-20, which have C-N rings, cluster in both low-temperature and high-temperature decomposition product spaces, suggesting similar decomposition patterns, which consistent with our previous findings [78]. DTTO and NTO, lacking the -$NO_2$ group, align more with the mid-to-high temperature space of the C-N ring containing HEMs of RDX, HMX, and CL-20. Phenyl ring HEMs (TNT, DNBF, HNS, TNB, TATB) also exhibit a high degree of overlap in chemical space, indicating similar structures and decomposition behaviors. In chain-like HEMs, PETN, NG, and BTTN, which contain multiple $NO_2$ groups (≥3), show better clustering, while FOX-7, with fewer $NO_2$ groups, predominantly appears in the mid-to-high temperature active space of chain-like HEMs after the $NO_2$ elimination reaction. For ionic salt HEMs, TAGN and TKX-50 cluster well, while ADN separates due to its $NH_4^+$ group and lack of carbon elements, unlike TAGN and TKX-50, which contain $NH_3OH^+$ and $NO_3^-$ groups.

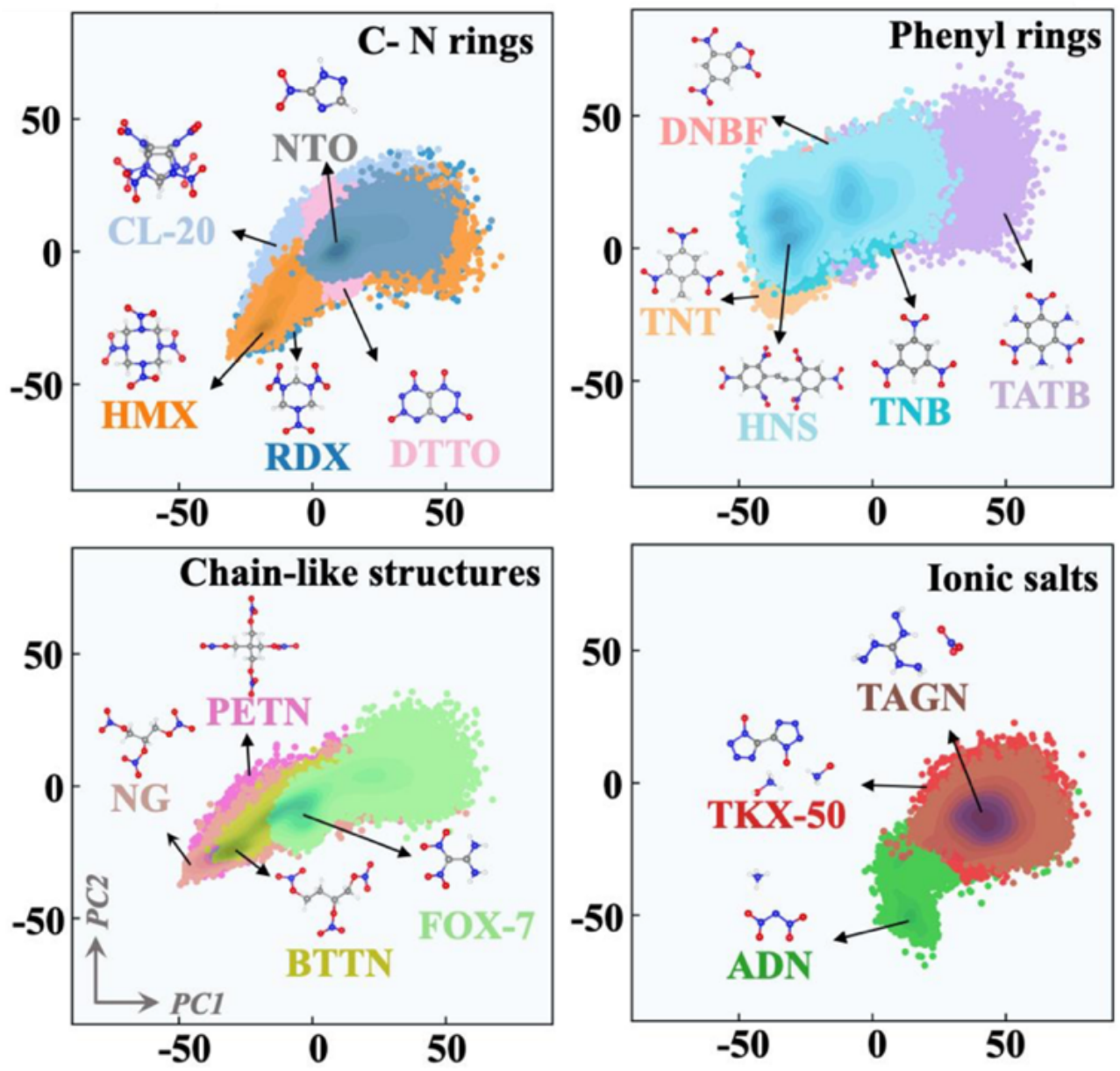

**Fig. 8** PCA visualization results of the configuration space of HEMs datasets with four different type structures.

PCA clustering and sample correlation analysis of the 20 HEMs dataset during NNP model training reveal the formation of active space for C, H, N, O-based HEMs and their chemical interactions. The results show that the general NNP model simplifies complex descriptors by learning the chemical structure space of similar species, supporting its design from first principles. This model enhances the understanding of complex chemical systems and offers a powerful tool for exploring chemical space and microscopic structural properties between molecules and condensed phases. Therefore, the general NNP model shows promising prospects as an indispensable and powerful method for next-generation molecular design and material optimization in chemical space modeling and microscopic mechanism prediction.

## 4 Conclusions

This work successfully develops an accurate and efficient general NNP model for C, H, N, O-based energetic materials (HEMs). Based on our previous pre-trained model (NNP-CHNO-2024) [42], this innovative approach combines a deep learning framework with the transfer learning method. By employing the transfer learning process and incorporating a small amount of new training data, the generalization ability and predictive accuracy of the model are significantly enhanced while reducing computational costs. The performance of NNP model was evaluated using a DFT database, showing that prediction errors for atomic energies and forces across a wide temperature range (300-4000K) were below 65.2 meV/atom and 0.684 eV/Å, respectively. For mechanical properties, the predictions of the model for cell parameters and equations of state were highly consistent with DFT results and outperformed the ReaxFF force field. In thermal decomposition, MD simulations for 20 HEMs showed that the NNP model accurately predicted the distribution of major decomposition products under heating conditions, aligning well with experimental data. This confirms that the model can describe atomic-scale crystal structural, mechanical, and decomposition properties of HEMs in both finite and extended systems with high precision. Notably, the success of this model stems from its ability to effectively capture and predict the chemical activity space of HEMs, identifying key atomic interactions and reaction mechanisms during thermal decomposition. By accurately representing the chemical activity space, the NNP model enhances our understanding of how HEMs react under extreme conditions, facilitating predictions of their behavior in diverse environments. This pioneering work promises to accelerate the development and application of HEMs, supporting the creation of future models for their complex microstructures and macroscopic properties.

**Acknowledgments**

This work is supported by the National Natural Science Foundation of China (Grant 52106130) and the State Key Laboratory of Explosion Science and Safety Protection (Grants ZDKT21-01, QNKT23-15). The authors also acknowledge the support from the Foundation of Science and Technology on Combustion and Explosion Laboratory. DC acknowledges the Science and Technology Innovation Program of Beijing Institute of Technology (2022CX01028).

## Declaration of competing interest

The authors declare that they have no known competing financial interests or personal relationships that could have appeared to influence the work reported in this paper.

# Supplementary Material (SM)

## A General Neural Network Potential for Energetic Materials with C, H, N, and O elements

**Mingjie Wen, Jiahe Han, Wenjuan Li, Xiaoya Chang, Qingzhao Chu*, Dongping Chen**

*State Key Laboratory of Explosion Science and Safety Protection, Beijing Institute of Technology, Beijing 100081, P. R. China*

In the supplementary materials, we present:

1. **Table S1** The types of HEMs involved in this NNP model.
2. **Fig. S1** The evaluation of energy (eV/atom) (a) and forces (eV/Å) (b) predicted by the pre-trained NNP model compared to DFT calculations.
3. **Fig. S2** MAEs of the energy (eV/atom) (a) and force (eV/Å) (b) predicted by the general NNP model (horizontal axis) and the DFT calculations (vertical axis).
4. **Fig. S3** Comparison of the MAE distributions (vertical axis) between the training set predictions obtained using the generalized NNP model and the DFT calculations for energy predictions (eV/atom) (a) and force predictions (eV/Å) (b).
5. **Table S2** Crystallographic parameters of all HEM crystals calculated using DFT, NNP, and ReaxFF methods.
6. **Fig. S4** EOS curves for all HEM crystals. "NNP" refers to the NNP model developed in this study. "DFT" computed at the PBE/DZVP-MOLOPT level using CP2K, and "ReaxFF" taken from the work of Liu et al. Dashed regions indicate data included in the training set.
7. **Fig. S5** Thermal decomposition simulation results of all HEM crystals heated from 300 K to 3000 K obtained using the general NNP model.
8. **Fig. S6** PCA visualization results of the single component and configuration space of the 20 HEMs dataset at 300 K during the NNP model training (a) and the sample correlation heatmap (b).
9. **Fig. S7** PCA visualization results of the single component and configuration space of the 20 HEMs dataset at 1500 K during the NNP model training (a) and the sample correlation heatmap (b).
10. **Fig. S8** PCA visualization results of the single component and configuration space of the 20 HEMs dataset at 3000 K during the NNP model training (a) and the sample correlation heatmap (b).

* Corresponding author, E-mail: chuqz@bit.edu.cn (Q. Z. Chu).

**Table S1** The types of HEMs involved in this NNP model.

| Abbreviation | Names | Abbreviation | Names |
|---|---|---|---|
| RDX | 1,3,5-trinitroperhydro-1,3,5-triazine | TAGN | 1,2,3-Triaminoguanidine nitrate |
| CL-20 | 2,4,6,8,10,12-hexanitrohexaazaisowurtzitane | NG | Nitroglycerin |
| HMX | 1,3,5,7-Tetranitro-1,3,5,7-tetrazocane | PETN | Pentaerythritol tetranitrate |
| TNT | 2-Methyl-1,3,5-trinitrobenzene | DTTO | Di-1,2,3,4-tetrazine Tetraoxides |
| ADN | Ammonium dinitramide | NTO | 3-Nitro-1,2,4-triazole-5-one |
| FOX-7 | 1,1-Diamino-2,2-dinitroethylene | TEX | 4,10-Dinitro-2,6,8,12-tetraoxa-4,10-diazawurtzitane |
| TKX-50 | Dihydroxylammonium 5,5′-bistetrazole-1,1′-diolate | BTTN | 1,2,4-Butanetriol trinitrate |
| DNBF | 4,6-dinitrobenzofuroxan | NC | Nitrocellulose |
| BTF | Benzotrifuroxan | TNB | 1,3,5-trinitrobenzene |
| TATB | 2,4,6-Trinitrobenzene-1,3,5-triamine | HNS | 2,2′,4,4′,6,6′-Hexanitrostilbene |

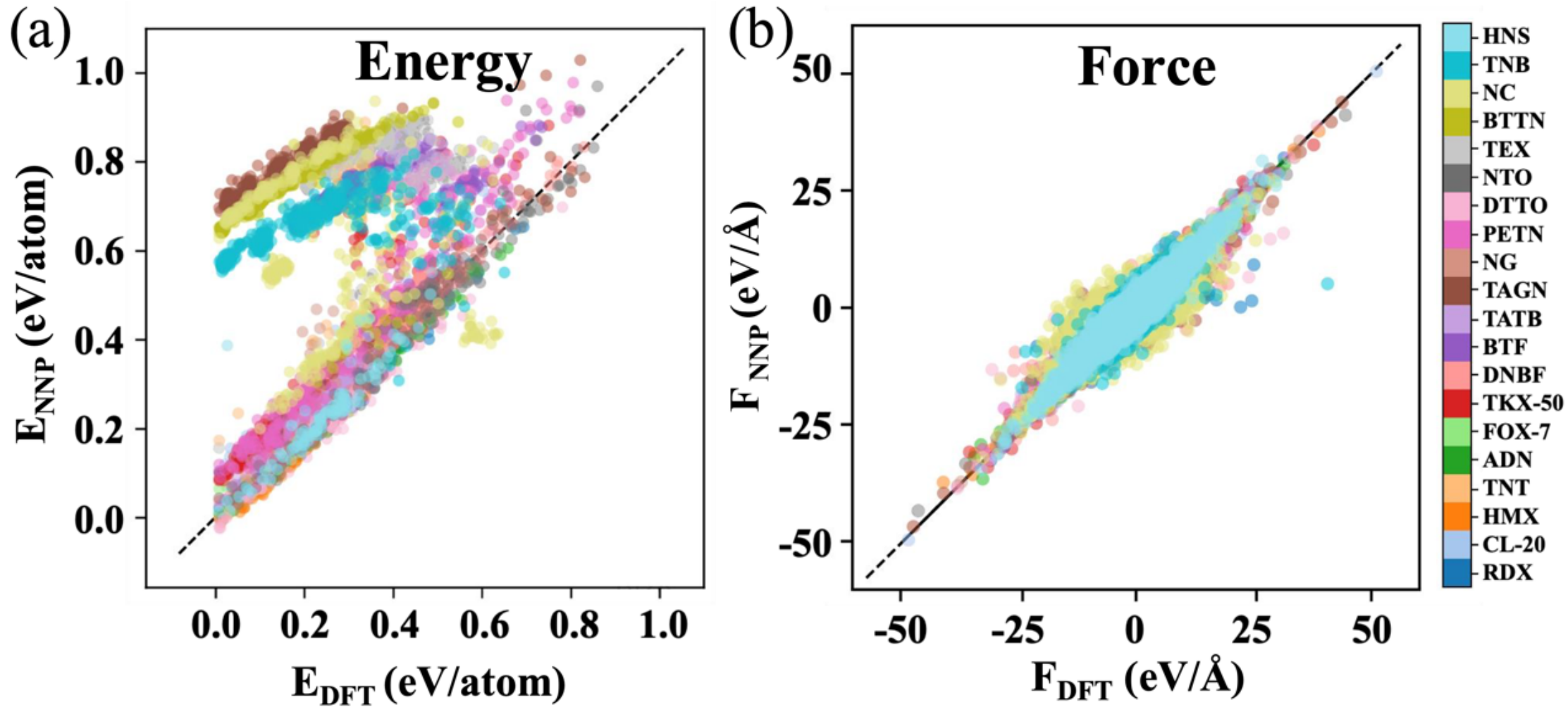

**Fig. S1** The evaluation of energy (eV/atom) (a) and forces (eV/Å) (b) predicted by the pre-trained NNP model [1] compared to DFT calculations.

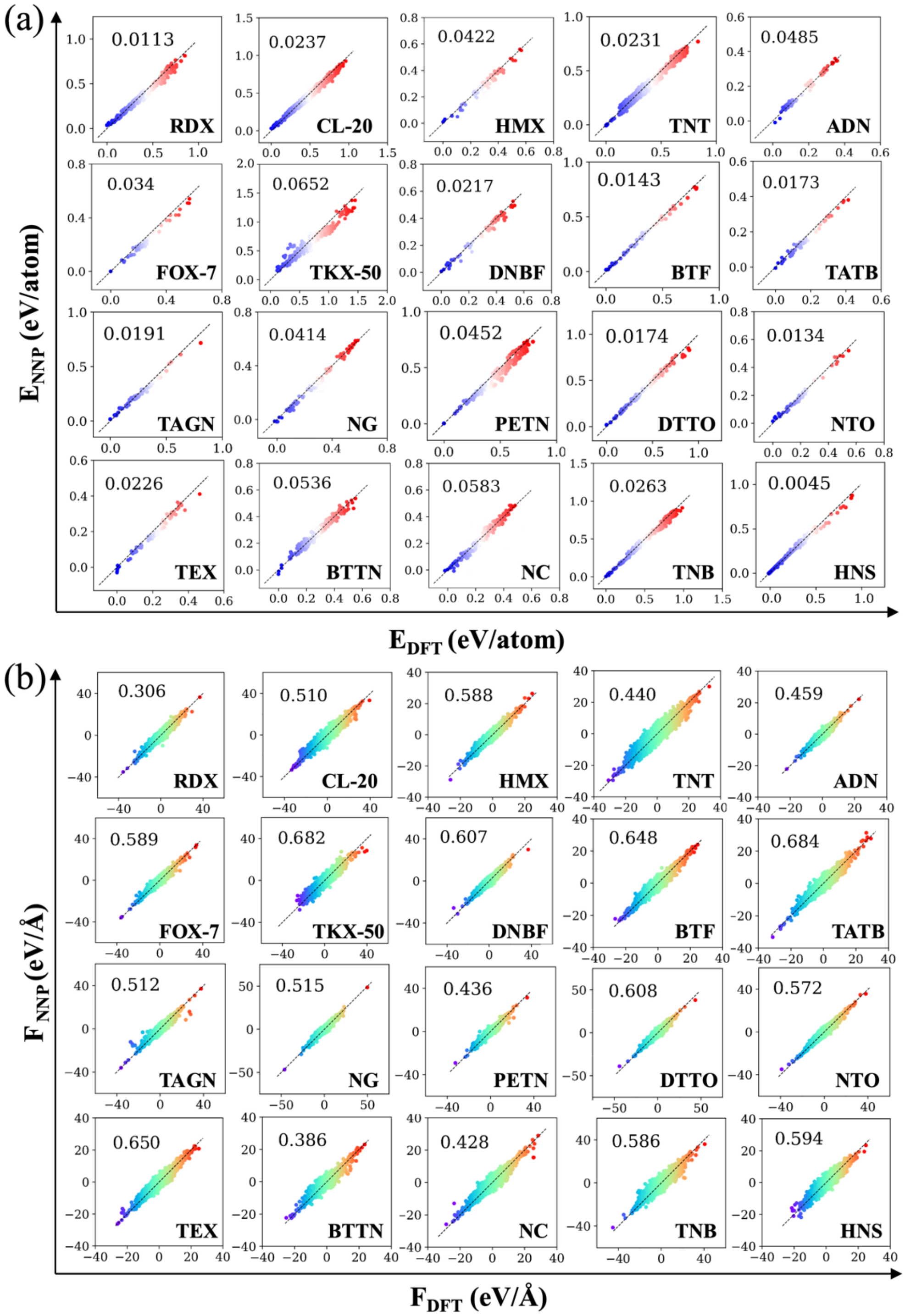

**Fig. S2** MAEs of the energy (eV/atom) (a) and force (eV/Å) (b) predicted by the general NNP model (horizontal axis) and the DFT calculations (vertical axis).

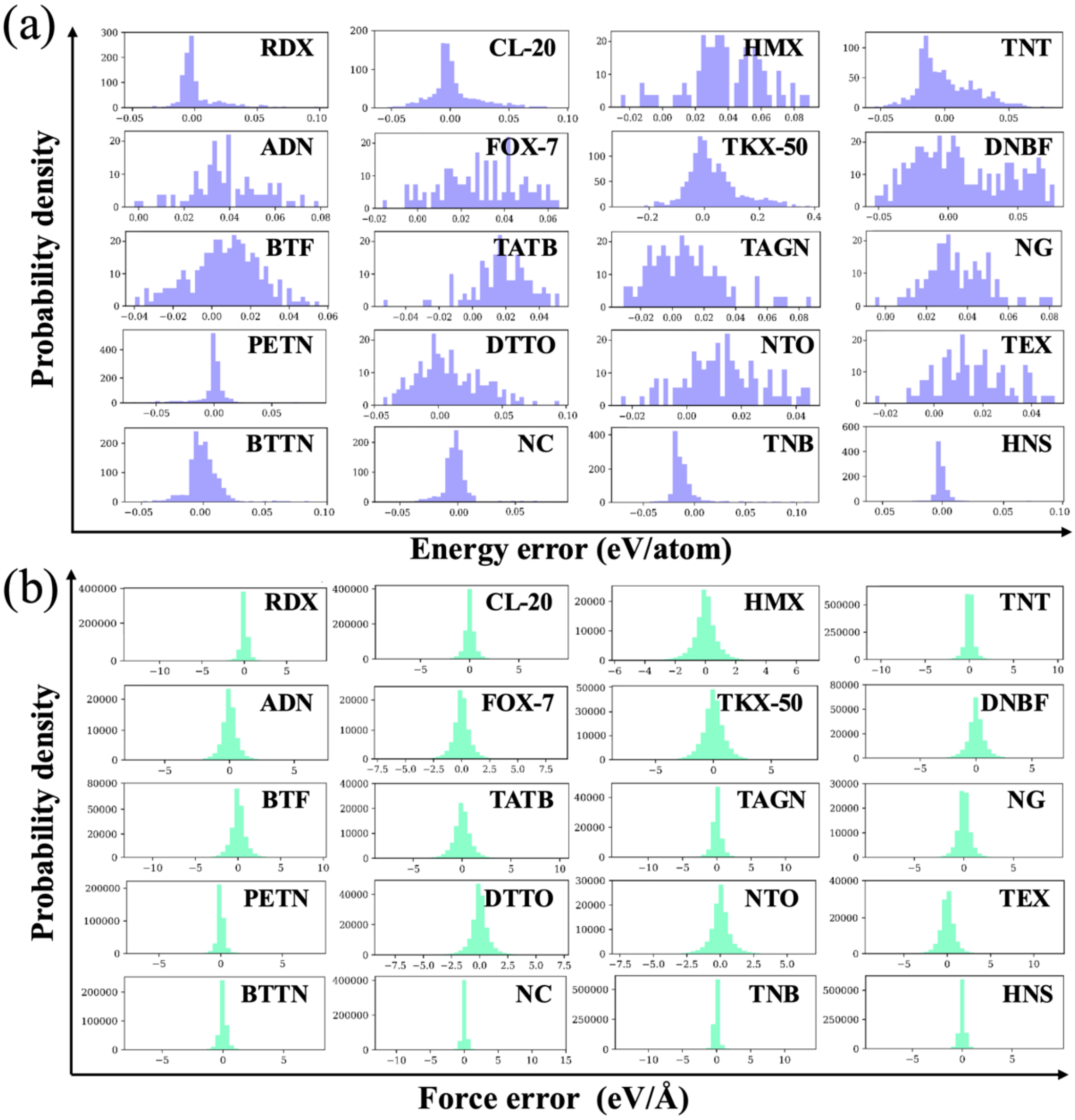

**Fig. S3** Comparison of the MAE distributions (vertical axis) between the training set predictions obtained using the generalized NNP model and the DFT calculations for energy predictions (eV/atom) (a) and force predictions (eV/Å) (b).

**Table S2** Crystallographic parameters of all HEM crystals calculated using DFT, NNP, and ReaxFF methods.[a]

| Parameter | | Experiment [b] | DFT [c] | NNP | ReaxFF [d] |
|---|---|---|---|---|---|
| RDX | a (Å) | 11.45 | 11.59 (1.18%) | 11.64 (1.64%), (0.45%) | 12.05 (5.20%), (3.97%) |
| | b (Å) | 10.62 | 10.74 (1.19%) | 10.79 (1.64%), (0.44%) | 11.34 (6.81%), (5.56%) |
| | c (Å) | 13.18 | 13.30 (0.90%) | 13.39 (1.64%), (0.73%) | 13.72 (4.12%), (3.19%) |
| | V (Å$^3$) | 1602.51 | 1655.52 (3.31%) | 1682.47 (4.99%), (1.63%) | 1874.80 (16.99%), (13.25%) |
| CL-20 | a (Å) | 12.58 | 12.74 (1.28%) | 12.95 (2.95%), (1.65%) | 13.41 (6.61%), (5.26%) |
| | b (Å) | 7.72 | 7.88 (2.05%) | 7.95 (2.95%), (0.89%) | 8.23 (6.58%), (4.44%) |
| | c (Å) | 13.17 | 13.32 (1.15%) | 13.56 (2.97%), (1.80%) | 14.04 (6.62%), (5.41%) |
| | V (Å$^3$) | 1279.1 | 1337.10 (4.53%) | 1397.23 (9.24%), (4.50%) | 1549.44 (21.14%), (15.88%) |
| HMX | a (Å) | 6.53 | 6.53 (0.02%) | 6.62 (1.40%), (1.38%) | 6.80 (4.15%), (4.13%) |
| | b (Å) | 10.99 | 10.92 (0.61%) | 11.09 (0.93%), (1.56%) | 11.38 (3.57%), (4.21%) |
| | c (Å) | 7.35 | 7.34 (0.07%) | 7.31 (0.48%), (0.41%) | 7.46 (1.56%), (1.63%) |
| | V (Å$^3$) | 514.2 | 510.49 (0.72%) | 521.02 (1.33%), (2.06%) | 564.74 (9.83%), (10.63%) |
| TNT | a (Å) | 14.99 | 15.21 (1.49%) | 15.30 (2.08%), (0.58%) | 16.08 (7.26%), (5.69%) |
| | b (Å) | 6.08 | 6.20 (2.04%) | 6.20 (2.08%), (0.04%) | 6.62 (8.94%), (6.76%) |
| | c (Å) | 20.02 | 20.08 (0.33%) | 20.43 (2.08%), (1.74%) | 21.12 (5.51%), (5.16%) |
| | V (Å$^3$) | 1823.55 | 1894.77 (3.91%) | 1939.78 (6.37%), (2.38%) | 2248.22 (23.29%), (18.65%) |
| ADN | a (Å) | 6.91 | 6.96 (0.67%) | 7.25 (4.86%), (4.17%) | 7.50 (8.48%), (7.76%) |
| | b (Å) | 11.79 | 11.55 (2.01%) | 12.48 (5.88%), (8.05%) | 12.92 (9.61%), (11.86%) |
| | c (Å) | 5.61 | 5.98 (6.52%) | 5.90 (5.09%), (1.34%) | 6.02 (7.23%), (0.67%) |
| | V (Å$^3$) | 450 | 472.42 (4.98%) | 518.69 (15.26%), (9.79%) | 576.24 (28.05%), (21.98%) |
| FOX7 | a (Å) | 6.94 | 7.06 (1.71%) | 7.18 (3.44%), (1.70%) | 7.22 (4.02%), (2.27%) |
| | b (Å) | 6.57 | 6.60 (0.47%) | 6.79 (3.36%), (2.88%) | 6.84 (4.13%), (3.64%) |
| | c (Å) | 11.32 | 11.40 (0.75%) | 11.70 (3.40%), (2.63%) | 11.77 (4.02%), (3.25%) |
| | V (Å$^3$) | 515.9 | 531.10 (2.95%) | 570.78 (10.64%), (7.47%) | 581.29 (12.67%), (9.45%) |
| TKX-50 | a (Å) | 5.49 | 5.44 (0.86%) | 5.67 (3.33%), (4.23%) | 5.76 (4.97%), (5.88%) |
| | b (Å) | 11.55 | 11.75 (1.76%) | 12.13 (5.05%), (3.23%) | 12.12 (4.96%), (3.15%) |
| | c (Å) | 6.48 | 6.32 (2.52%) | 6.67 (2.88%), (5.54%) | 6.78 (4.58%), (7.28%) |
| | V (Å$^3$) | 408.97 | 402.05 (1.69%) | 456.74 (11.68%), (13.60%) | 471.39 (15.26%), (17.25%) |
| DNBF | a (Å) | 7.41 | 7.47 (0.84%) | 7.60 (2.59%), (1.74%) | 7.75 (4.62%), (3.75%) |
| | b (Å) | 6.19 | 6.22 (0.57%) | 6.34 (2.51%), (1.93%) | 6.47 (4.61%), (4.02%) |
| | c (Å) | 9.8 | 9.81 (0.14%) | 9.76 (0.37%), (0.51%) | 9.74 (0.57%), (0.71%) |
| | V (Å$^3$) | 426.9 | 433.04 (1.44%) | 438.42 (2.70%), (1.24%) | 465.62 (9.07%), (7.52%) |
| BTF | a (Å) | 6.92 | 6.98 (0.82%) | 7.28 (5.16%), (4.30%) | 7.23 (4.43%), (3.58%) |
| | b (Å) | 19.52 | 19.54 (0.12%) | 20.41 (4.58%), (4.45%) | 20.38 (4.43%), (4.30%) |
| | c (Å) | 6.52 | 6.82 (4.63%) | 6.90 (5.86%), (1.17%) | 6.81 (4.48%), (0.15%) |
| | V (Å$^3$) | 880.64 | 929.50 (5.55%) | 1025.24 (16.42%), (10.30%) | 1002.85 (13.88%), (7.89%) |
| TATB | a (Å) | 9.01 | 9.03 (0.22%) | 9.18 (1.89%), (1.66%) | 9.33 (3.55%), (3.32%) |
| | b (Å) | 9.03 | 9.04 (0.13%) | 9.38 (3.90%), (3.76%) | 10.01 (10.88%), (10.73%) |
| | c (Å) | 6.81 | 6.61 (2.97%) | 6.62 (2.82%), (0.15%) | 6.85 (0.56%), (3.63%) |
| | V (Å$^3$) | 442.52 | 428.68 (3.13%) | 458.91 (3.70%), (7.05%) | 528.73 (19.48%), (23.34%) |
| TAGN | a (Å) | 8.39 | 8.35 (0.46%) | 8.76 (4.42%), (4.91%) | 9.23 (10.03%), (10.54%) |
| | b (Å) | 12.68 | 12.66 (0.19%) | 13.24 (4.38%), (4.58%) | 13.55 (6.83%), (7.03%) |
| | c (Å) | 6.54 | 6.78 (3.62%) | 6.83 (4.39%), (0.74%) | 7.20 (10.04%), (6.19%) |
| | V (Å$^3$) | 696.22 | 717.37 (3.04%) | 791.57 (13.70%), (10.34%) | 900.48 (29.34%), (25.53%) |
| NG | a (Å) | 8.9 | 9.06 (1.80%) | 9.21 (3.48%), (1.66%) | 9.38 (5.39%), (3.53%) |
| | b (Å) | 13.61 | 13.81 (1.48%) | 14.08 (3.47%), (1.96%) | 14.34 (5.38%), (3.84%) |
| | c (Å) | 6.76 | 6.92 (2.34%) | 7.00 (3.52%), (1.16%) | 7.13 (5.44%), (3.03%) |
| | V (Å$^3$) | 818.95 | 865.31 (5.66%) | 907.38 (10.80%), (4.86%) | 958.92 (17.09%), (10.82%) |
| PETN | a (Å) | 13.29 | 13.37 (0.60%) | 13.49 (1.50%), (0.90%) | 13.77 (3.61%), (2.99%) |
| | b (Å) | 13.49 | 13.58 (0.67%) | 13.70 (1.56%), (0.88%) | 13.98 (3.63%), (2.95%) |
| | c (Å) | 6.83 | 6.90 (1.02%) | 6.93 (1.46%), (0.43%) | 7.08 (3.66%), (2.61%) |
| | V (Å$^3$) | 1224.5 | 1252.60 (2.29%) | 1281.65 (4.67%), (2.32%) | 1361.93 (11.22%), (8.73%) |
| DTTO | a (Å) | 25.54 | 25.64 (0.39%) | 25.86 (1.25%), (0.86%) | 26.83 (5.05%), (4.64%) |
| | b (Å) | 5.43 | 5.45 (0.37%) | 5.50 (1.29%), (0.92%) | 5.71 (5.16%), (4.77%) |
| | c (Å) | 9.69 | 10.05 (3.77%) | 9.81 (1.29%), (2.39%) | 10.17 (5.01%), (1.19%) |
| | V (Å$^3$) | 1343.88 | 1404.37 (4.50%) | 1395.35 (3.83%), (0.64%) | 1557.24 (15.88%), (10.89%) |
| NTO | a (Å) | 9.31 | 9.42 (1.15%) | 9.50 (2.01%), (0.85%) | 9.86 (5.87%), (4.67%) |
| | b (Å) | 5.45 | 5.58 (2.46%) | 5.63 (3.38%), (0.90%) | 5.84 (7.24%), (4.66%) |
| | c (Å) | 9.03 | 9.12 (1.04%) | 9.12 (1.04%), (0.00%) | 9.46 (4.81%), (3.73%) |
| | V (Å$^3$) | 448.64 | 470.86 (4.95%) | 479.12 (6.79%), (1.75%) | 536.04 (19.48%), (13.84%) |
| TEX | a (Å) | 6.84 | 6.97 (1.96%) | 7.03 (2.84%), (0.86%) | 7.18 (5.03%), (3.01%) |
| | b (Å) | 7.64 | 7.69 (0.65%) | 7.73 (1.17%), (0.52%) | 7.88 (3.14%), (2.47%) |
| | c (Å) | 8.78 | 8.92 (1.64%) | 8.79 (0.15%), (1.46%) | 8.86 (0.95%), (0.67%) |
| | V (Å$^3$) | 433.64 | 452.32 (4.31%) | 446.06 (2.86%), (1.38%) | 475.62 (9.68%), (5.15%) |
| BTTN | a (Å) | | 17.11 | 17.55 (2.57%) | 17.89 (4.56%) |
| | b (Å) | | 11.56 | 11.73 (1.47%) | 11.88 (2.77%) |
| | c (Å) | | 11.55 | 11.59 (0.35%) | 11.70 (1.30%) |
| | V (Å$^3$) | | 2284.49 | 2385.93 (4.44%) | 2486.64 (8.85%) |
| NC | a (Å) | 10.4 | 10.41 (0.13%) | 10.45 (0.51%), (0.38%) | 10.59 (1.83%), (1.69%) |
| | b (Å) | 6.1 | 6.11 (0.12%) | 6.13 (0.45%), (0.33%) | 6.42 (5.29%), (5.16%) |
| | c (Å) | 5.26 | 5.26 (0.07%) | 5.29 (0.70%), (0.63%) | 5.42 (3.11%), (3.04%) |
| | V (Å$^3$) | 333.3 | 334.42 (0.34%) | 338.91 (1.68%), (1.34%) | 368.49 (10.56%), (10.19%) |
| TNB | a (Å) | 12.59 | 12.86 (2.17%) | 12.87 (2.25%), (0.08%) | 13.18 (4.71%), (2.49%) |
| | b (Å) | 9.68 | 9.89 (2.13%) | 9.90 (2.23%), (0.10%) | 10.14 (4.71%), (2.53%) |
| | c (Å) | 26.86 | 27.01 (0.56%) | 27.47 (2.27%), (1.70%) | 28.12 (4.69%), (4.11%) |
| | V (Å$^3$) | 3274 | 3433.26 (4.86%) | 3502.88 (6.99%), (2.03%) | 3756.76 (14.75%), (9.42%) |
| HNS | a (Å) | 22.08 | 22.33 (1.12%) | 22.47 (1.75%), (0.63%) | 22.65 (2.57%), (1.43%) |
| | b (Å) | 5.55 | 5.57 (0.29%) | 5.66 (1.91%), (1.62%) | 5.92 (6.59%), (6.28%) |
| | c (Å) | 14.63 | 14.65 (0.11%) | 14.67 (0.25%), (0.14%) | 15.06 (2.91%), (2.80%) |
| | V (Å$^3$) | 1702.59 | 1729.88 (1.60%) | 1773.48 (4.16%), (2.52%) | 1927.11 (13.19%), (11.40%) |

[a] The values in parentheses represent the deviations between the calculated results and experimental values, expressed as a percentage. The values in black parentheses denote the absolute errors between the current results and experimental data, while the values in blue parentheses denote the absolute errors between the current results and DFT calculation results.

[b] Experimental values are taken from the CCDC Crystallographic database via https://www.ccdc.cam.ac.uk/.

[c] The DFT calculations are computed at the PBE/DZVP-MOLOPT level using CP2K.

[d] The ReaxFF is taken from the work of Liu et al [2].

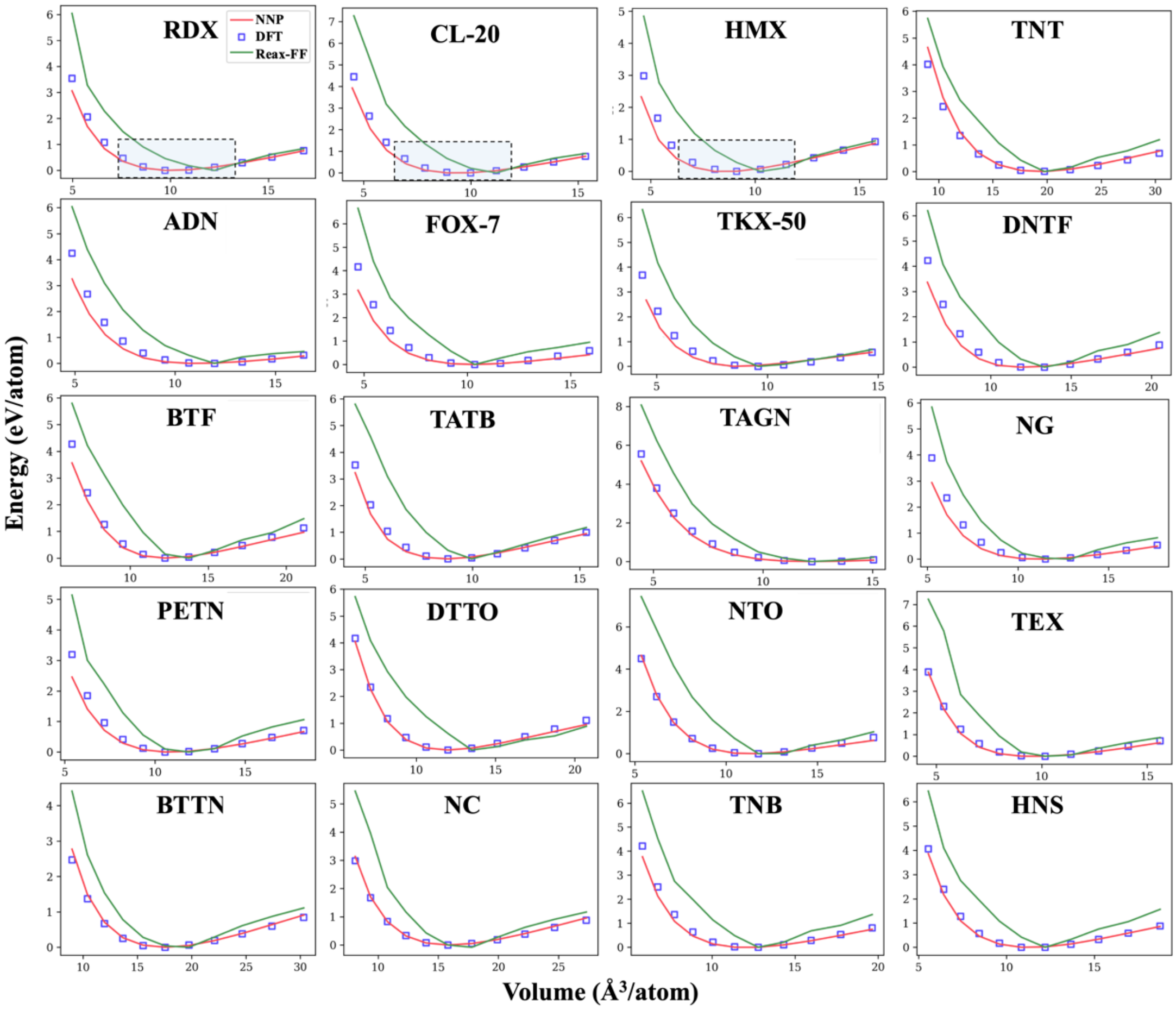

**Fig. S4** EOS curves for all HEM crystals. "NNP" refers to the NNP model developed in this study. "DFT" computed at the PBE/DZVP-MOLOPT level using CP2K [3], and "ReaxFF" taken from the work of Liu et al. [2]. Dashed regions indicate data included in the training set.

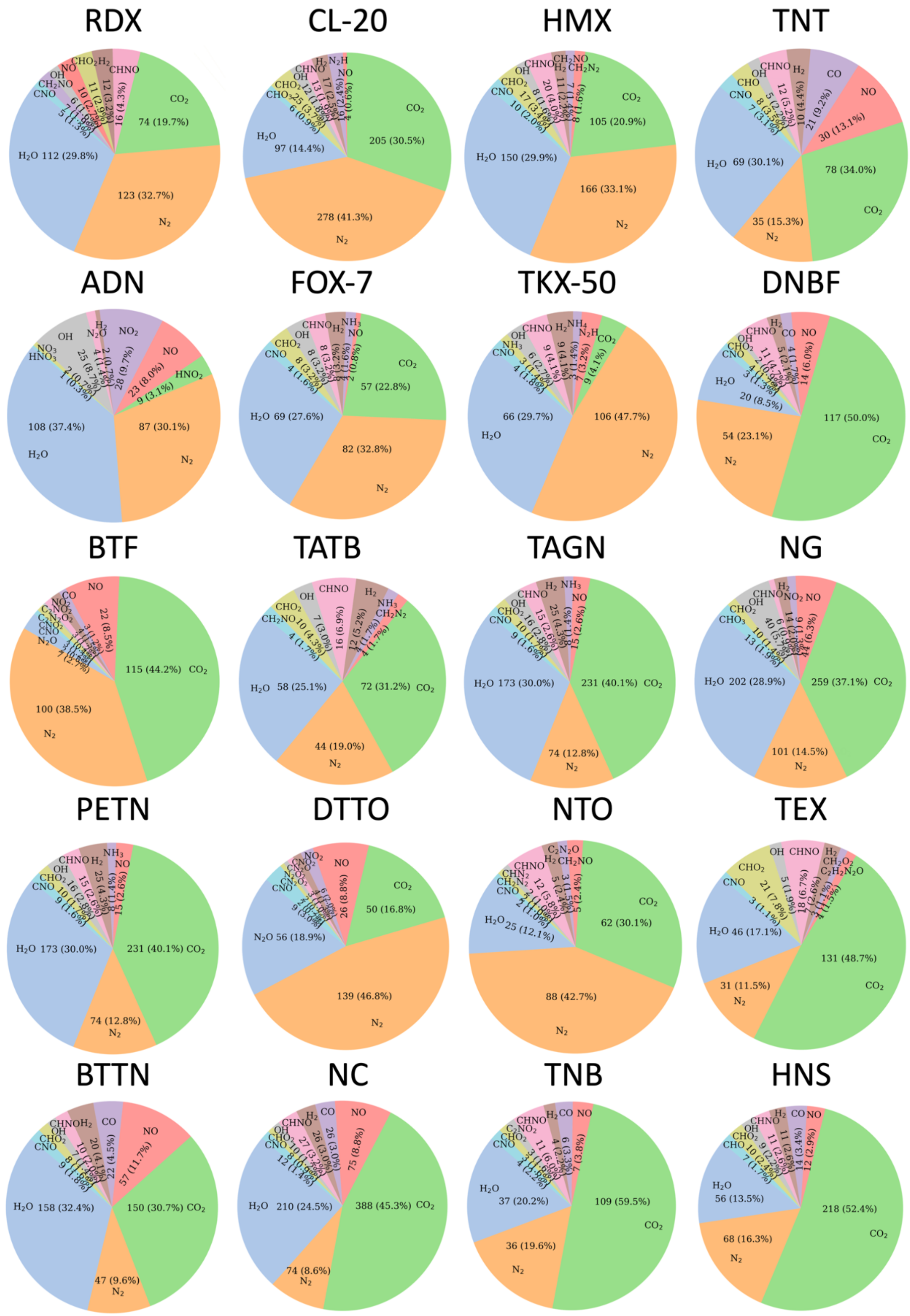

**Fig. S5** Thermal decomposition simulation results of all HEM crystals heated from 300 K to 3000 K obtained using the general NNP model.

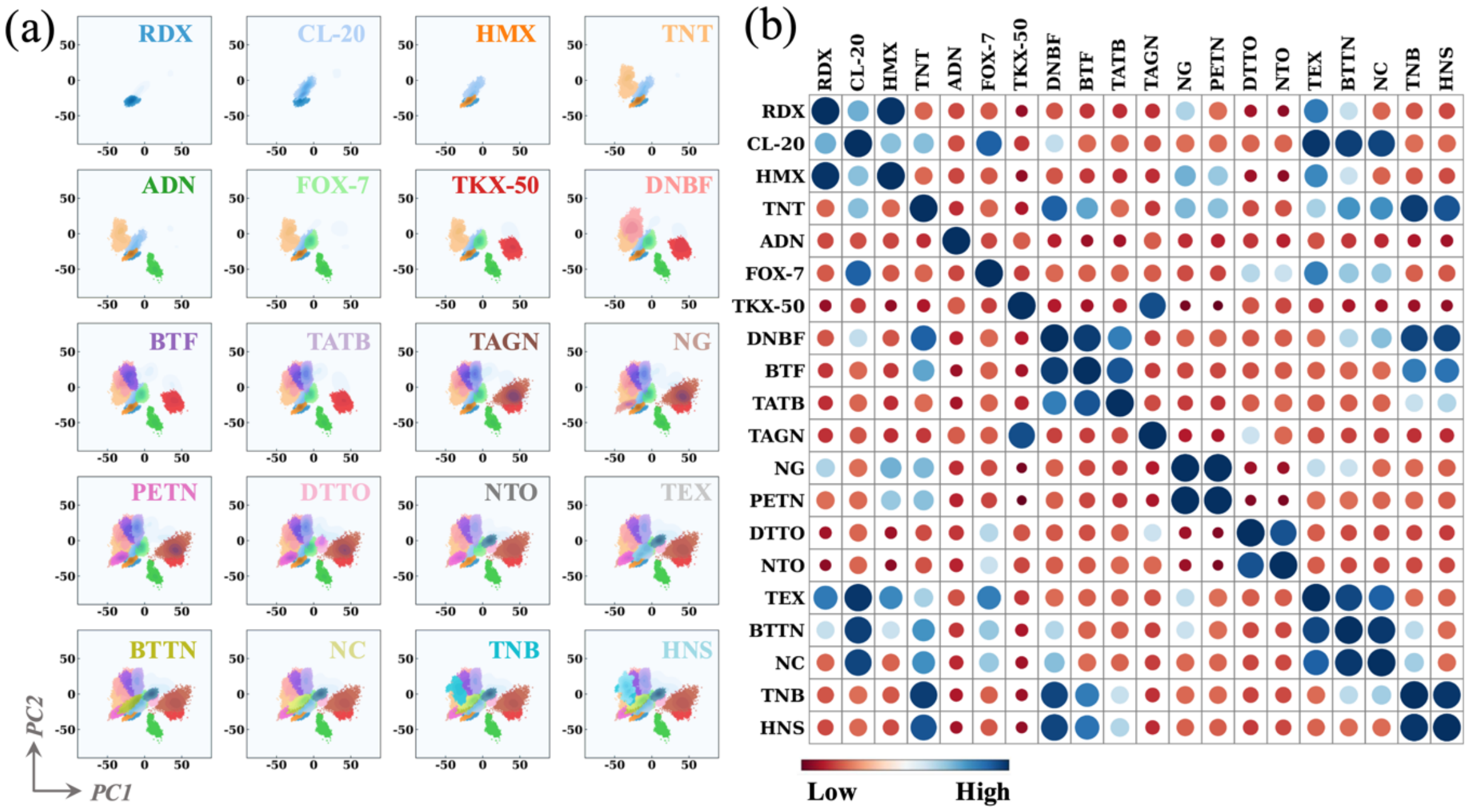

**Fig. S6** PCA visualization results of the single component and configuration space of the 20 HEMs dataset at 300 K during the NNP model training (a) and the sample correlation heatmap (b).

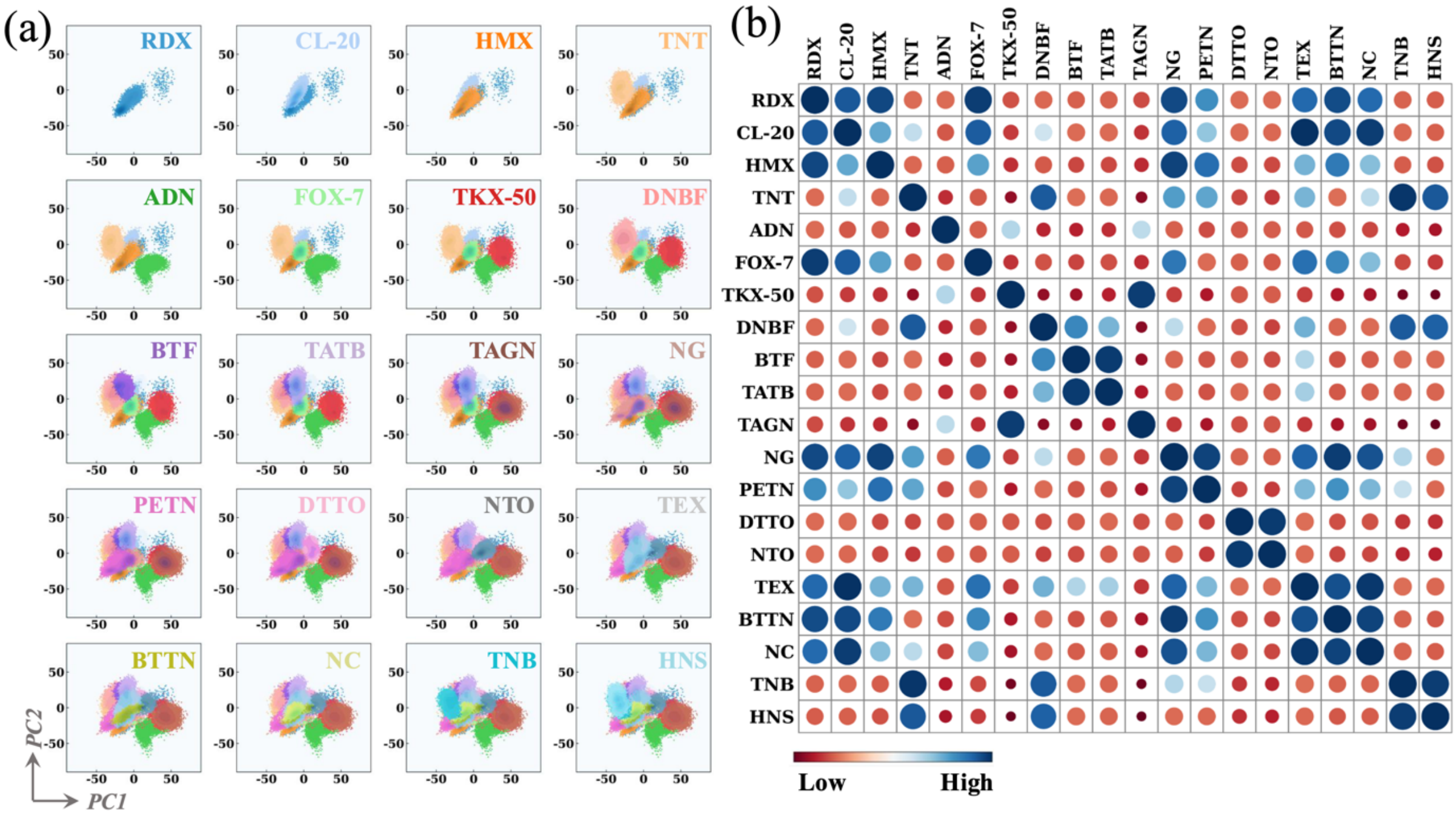

**Fig. S7** PCA visualization results of the single component and configuration space of the 20 HEMs dataset at 1500 K during the NNP model training (a) and the sample correlation heatmap (b).

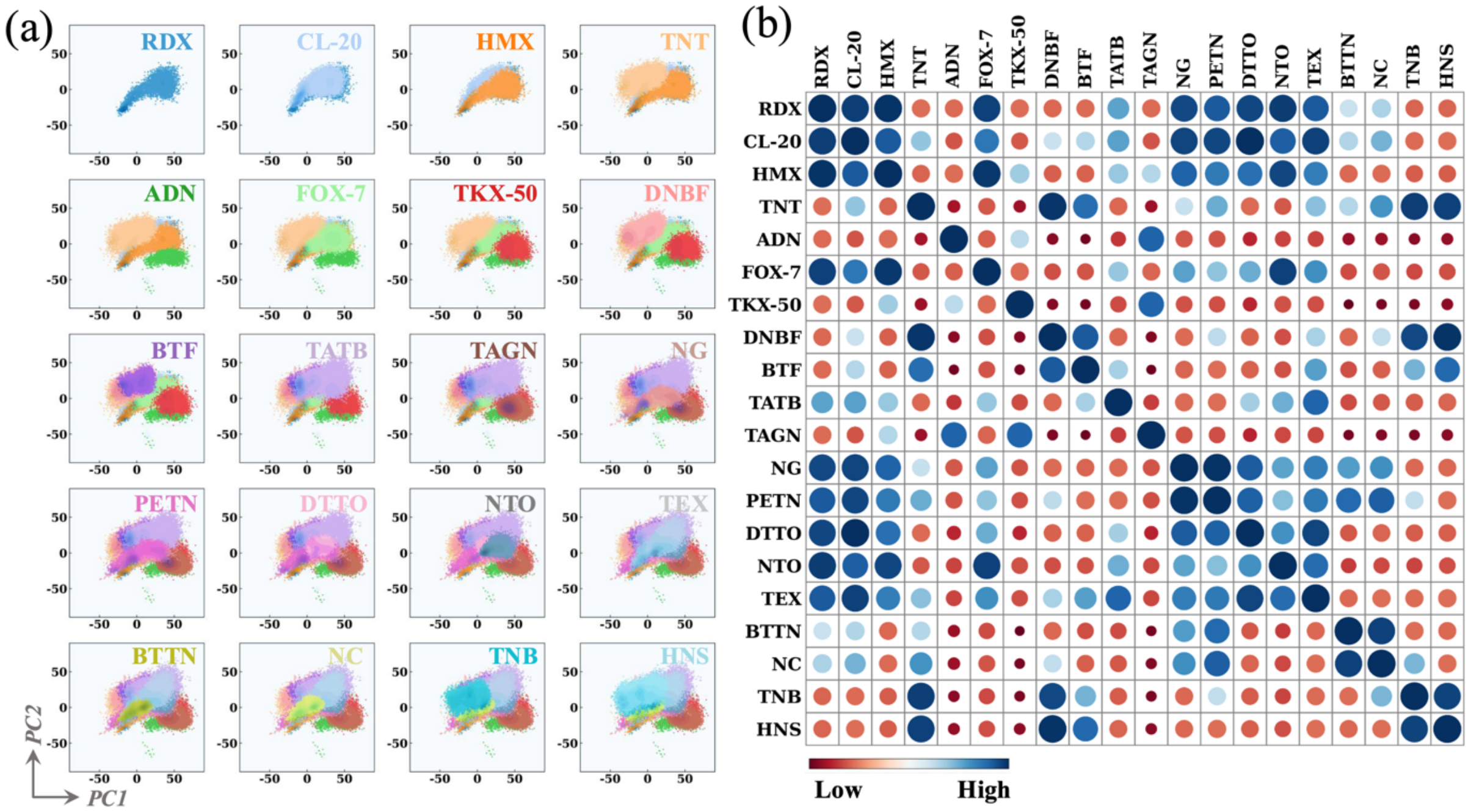

**Fig. S8** PCA visualization results of the single component and configuration space of the 20 HEMs dataset at 3000 K during the NNP model training (a) and the sample correlation heatmap (b).